\documentclass[onecolumn]{IEEEtran}

\usepackage[mathscr]{eucal}
\usepackage[cmex10]{amsmath}
\usepackage{epsfig,epsf,psfrag}
\usepackage{amssymb,amsmath,amsthm,amsfonts,latexsym}
\usepackage{amsmath,graphicx,bm,xcolor,url}
\usepackage[caption=false]{subfig} 
\usepackage{fixltx2e}
\usepackage{array}
\usepackage{verbatim}
\usepackage{bm}
\usepackage{algorithmic, cite}
\usepackage{algorithm}
\usepackage{verbatim}
\usepackage{textcomp}
\usepackage{mathrsfs}
\usepackage{multirow}
\usepackage{epstopdf}

\catcode`~=11 \def\UrlSpecials{\do\~{\kern -.15em\lower .7ex\hbox{~}\kern .04em}} \catcode`~=13 

\allowdisplaybreaks[1]
 
\newcommand{\nn}{\nonumber}

\newcommand{\calA}{\mathcal{A}}

\newcommand{\calF}{\mathcal{F}}

\newcommand{\calH}{\mathcal{H}}

\newcommand{\calM}{\mathcal{M}}

\newcommand{\calS}{\mathcal{S}}

\newcommand{\calW}{\mathcal{W}}
\newcommand{\calX}{\mathcal{X}}
\newcommand{\calY}{\mathcal{Y}}
\newcommand{\calZ}{\mathcal{Z}}

\newcommand{\rma}{\mathrm{a}}

\newcommand{\rmd}{\mathrm{d}}

\newcommand{\rme}{\mathrm{e}}

\newcommand{\rmm}{\mathrm{m}}

\newcommand{\rmx}{\mathrm{x}}

\newcommand{\bbE}{\mathbb{E}}

\newcommand{\bbN}{\mathbb{N}}

\newcommand{\bbP}{\mathbb{P}}

\newcommand{\bbR}{\mathbb{R}}

\DeclareMathAlphabet{\mathbsf}{OT1}{cmss}{bx}{n}
\DeclareMathAlphabet{\mathssf}{OT1}{cmss}{m}{sl}

\newcommand{\rvE}{\mathsf{E}}

\newcommand{\rvH}{\mathsf{H}}

\newcommand{\rvP}{\mathsf{P}}

\DeclareSymbolFont{bsfletters}{OT1}{cmss}{bx}{n}  
\DeclareSymbolFont{ssfletters}{OT1}{cmss}{m}{n}
\DeclareMathSymbol{\bsfGamma}{0}{bsfletters}{'000}
\DeclareMathSymbol{\ssfGamma}{0}{ssfletters}{'000}
\DeclareMathSymbol{\bsfDelta}{0}{bsfletters}{'001}
\DeclareMathSymbol{\ssfDelta}{0}{ssfletters}{'001}
\DeclareMathSymbol{\bsfTheta}{0}{bsfletters}{'002}
\DeclareMathSymbol{\ssfTheta}{0}{ssfletters}{'002}
\DeclareMathSymbol{\bsfLambda}{0}{bsfletters}{'003}
\DeclareMathSymbol{\ssfLambda}{0}{ssfletters}{'003}
\DeclareMathSymbol{\bsfXi}{0}{bsfletters}{'004}
\DeclareMathSymbol{\ssfXi}{0}{ssfletters}{'004}
\DeclareMathSymbol{\bsfPi}{0}{bsfletters}{'005}
\DeclareMathSymbol{\ssfPi}{0}{ssfletters}{'005}
\DeclareMathSymbol{\bsfSigma}{0}{bsfletters}{'006}
\DeclareMathSymbol{\ssfSigma}{0}{ssfletters}{'006}
\DeclareMathSymbol{\bsfUpsilon}{0}{bsfletters}{'007}
\DeclareMathSymbol{\ssfUpsilon}{0}{ssfletters}{'007}
\DeclareMathSymbol{\bsfPhi}{0}{bsfletters}{'010}
\DeclareMathSymbol{\ssfPhi}{0}{ssfletters}{'010}
\DeclareMathSymbol{\bsfPsi}{0}{bsfletters}{'011}
\DeclareMathSymbol{\ssfPsi}{0}{ssfletters}{'011}
\DeclareMathSymbol{\bsfOmega}{0}{bsfletters}{'012}
\DeclareMathSymbol{\ssfOmega}{0}{ssfletters}{'012}

\newcommand{\tilf}{\tilde{f}}

\newcommand{\eps}{\varepsilon}

\DeclareMathOperator*{\argmax}{arg\,max}

\newcommand{\bone}{\mathbf{1}}

\newtheorem{theorem}{Theorem} 
\newtheorem{lemma}{Lemma}

\newtheorem{proposition}{Proposition}

\newtheorem{definition}{Definition}

\newtheorem{remark}{Remark}

\newcommand{\qednew}{\nobreak \ifvmode \relax \else
      \ifdim\lastskip<1.5em \hskip-\lastskip
      \hskip1.5em plus0em minus0.5em \fi \nobreak
      \vrule height0.75em width0.5em depth0.25em\fi}

\usepackage{xspace}
\usepackage[ colorlinks = true,
             linkcolor = blue,
             urlcolor  = blue,
             citecolor = red,
             anchorcolor = green,
]{hyperref}

\newcommand{\red}[1]{\textcolor{black}{#1}} 

\usepackage{cite}
\allowdisplaybreaks[2]
\begin{document} 

\title{Moderate Deviation  Asymptotics for Variable-Length Codes with Feedback} 

\author{Lan V.\ Truong  $\,$  and $\,$
        Vincent Y.~F.~Tan
\thanks{The authors are with the  Department of Electrical and Computer Engineering, National University of Singapore (NUS). V.~Y.~F.~Tan is also with the Department of Mathematics, NUS. Emails: \url{lantruong@u.nus.edu};  \url{vtan@nus.edu.sg}} 
}

\maketitle
\begin{abstract}
We consider   data transmission across discrete memoryless channels (DMCs) using  variable-length codes with feedback. We consider the family of such codes whose rates are $\rho_N$ below the channel capacity $C$,  where  $\rho_N$ is  a positive sequence that  tends to zero slower than the reciprocal of the square root of the expectation of the (random) blocklength $N$.  This is known as the moderate deviations regime and we establish the optimal  moderate deviations constant.  We show that in this scenario,  the error probability decays sub-exponentially with speed $\exp(-(B/C)N\rho_N)$, where $B$ is the maximum relative entropy between output distributions of the DMC. 
\end{abstract}   
\section{Introduction}
Shannon showed~\cite{Sha56} that feedback does not increase the capacity of memoryless channels. However, feedback has many practical advantages in various communication settings, including simplifying coding schemes~\cite{SK66, ShayevitzF} and dramatically minimizing the error probability at finite blocklengths~\cite{Burnashev1976, Yury2011}.  This paper focuses on the simplest channel model---the discrete memoryless channel (DMC). We consider the scenario in which the length of the code is a random variable (in particular, a stopping time of the filtration generated by the sequence of channel outputs) and full feedback  is available at the encoder. Different from previous works which consider the error exponents (large deviations) regime~\cite{Burnashev1976} and the fixed-error (second-order) regime~\cite{Yury2011}, we analyze the performance of codes whose transmission rates are $\rho_N$ below the channel capacity $C$,  where  $\rho_N$ is  a positive sequence that  tends to zero slower than the reciprocal of the square root of the expectation of the (random) blocklength $N$. This is known as the {\em moderate deviations regime}~\cite{altug14b, pol10e}. We derive a tight result for the {\em moderate deviations constant}; this is defined precisely in Section~\ref{sec:dmc_bc}.
\subsection{Related Works} \label{sec:related}
 Burnashev, in a seminal work~\cite{Burnashev1976}, proposed a communication model  for DMCs with feedback where the  blocklength $\tau\in\bbN$ is a random variable whose expectation is over bounded by some positive real number $N\in\bbR_+$. He demonstrated that the {\em reliability function} or {\em optimal error exponent}  for the DMC  with feedback improves dramatically over the no feedback case and the case where the blocklength is deterministic. This  class of codes is known as {\em variable-length  codes with feedback}. In fact, the reliability function of a DMC with variable-length feedback admits the particularly simple expression
\begin{align}
\rvE(R)= B \left(1-\frac{R}{C}\right),\quad\forall R\in [0,C],\label{eqn:burn}
\end{align} 
where $C$ is the capacity of the DMC and $B$ (usually written as $C_1$ in the literature) is  the relative entropy between conditional output distributions of the two most ``most distinguisable'' channel input symbols~\cite{Burnashev1976}. Yamamoto and Itoh~\cite{YamamotoItoh1979} proposed a simple and conceptually important two-phase coding scheme that achieves the reliability function in \eqref{eqn:burn}. Burnashev~\cite{Burnashev80} later extended the ideas in~\cite{Burnashev1976} to be amenable to the more general problem of sequential hypothesis testing. In particular, he studied the minimum expected number of observations (transmissions) to attain some level of reliability and found the  reliability function for large  class of single-user channels (beyond DMCs), including the Gaussian channel~\cite{Burnashev80}. Berlin {\em et al.}~\cite{Berlin2009a} provided a simple converse proof for Burnashev's reliability function~\cite{Burnashev1976} that parallels Yamamoto and Itoh's two-phase achievability scheme. Nakibo\u{g}lu and Gallager~\cite{Nak08} investigated variable-length coding schemes for (not necessarily discrete) memoryless channels with   variable-length feedback  and with cost constraints and established the reliability function.   
 Mahajan and Tatikonda~\cite{Mahajan12} considered the variable-length case for   compound channels~\cite{Blackwell59} and established inner and outer bounds on the so-called error exponent region. Tchamkerten  and Telatar,  in a series of elegant works~\cite{Tchamkerten05, Tchamkerten06, Tchamkerten06a}, considered conditions in which one can achieve Burnashev's exponent in \eqref{eqn:burn} universally, i.e., without precise knowledge of the DMC.   

For the above-mentioned works, we assume that the transmission rate is a fixed value below the channel capacity. It is natural to ask what happens in other regimes. One of the other regimes that has gained interest recently is the {\em second-order regime} in which the average error probability of the code $\epsilon>0$ is {\em non-vanishing}. Polyanskiy, Poor and Verd\'u~\cite{Yury2011} showed that $\epsilon$-capacity is enhanced by a factor of $\frac{1}{1-\epsilon}$ and optimum codes approach the $\epsilon$-capacity   rapidly---at a rate of $O\big(\frac{\ln N}{N} \big)$. Kostina, Polyanskiy, and Verd\'u~\cite{kost17} recently extended this framework to the joint source-channel coding scenario.


We analyze variable-length codes that operate in a regime that is ``sandwiched between'' the error exponent and second-order regimes so the codes have vanishing error probabilities {\em and} approach capacity as the lengths of the codes grow. This is known as the {\em moderate deviations regime}. Here,  we   mention a few existing works on moderate deviations analysis in information theory. Chen {\em et al.}~\cite{ChenHe2007e} and He {\em et al.}~\cite{HeMontano2009a}  studied the moderate deviation asymptotics for fixed-to-variable length source coding with decoder side information and cyclic symmetric channels.  Altu\u{g} and Wagner~\cite{altug14b} established the moderate deviations constant for DMCs by considering the behavior of the random coding and sphere packing exponents near capacity. Polyanksiy and Verd\'u~\cite{pol10e} provided a different derivation using the information spectrum method and they also considered  Gaussian channels.  In~\cite{altug14b}  and~\cite{pol10e}, it is shown that for a backoff from capacity of $\rho_n>0$ (where $\rho_n$ decays slower than the reciprocal of square root of the blocklength $n$), the error probability decays sub-exponentially   with speed $\exp(- n\rho_n^2 / (2V))$ where $V$ is the dispersion of the channel. Tan~\cite{Tan2012e} and    Altu\u{g}, Wagner and Kontoyiannis~\cite{AWK13} considered moderate deviations for lossy and lossless source coding respectively. 
Alt\u{u}g, Poor and Verd\'u~\cite{AltugPoorVerdu2015} studied the moderate deviations behavior for fixed-length  channel codes with  feedback.  The authors showed that, under some conditions on   DMCs~\cite[Corollary~1]{AltugPoorVerdu2015}, the moderate deviations constant $1/(2V)$ remains unchanged. In all works on moderate deviation asymptotics for fixed-length  block codes, the error probability scales as $\exp(- \Theta(n\rho_n^2))$. 

\subsection{Main Contributions}

We show that for variable-length codes with feedback in the moderate deviations regime, the error probability scales as  $\exp(- \Theta(N\rho_N))$, where $N$ is the expectation of the (random) blocklength and $\rho_N=\omega(1/\sqrt{N})$.  Moreover the implied constant in the $\Theta(\cdot)$ notation, known as the {\em optimal moderate deviations constant}, is $B/C$.   This is not overly surprising in light of Burnashev's result in~\eqref{eqn:burn} because if we take $R$ therein to be $R=C-\rho_N$, we obtain
\begin{equation}
\rvE(C-\rho_N) = B\left( 1-\frac{C-\rho_N}{C}\right)=\frac{B}{C} \cdot \rho_N.\label{eqn:heur}
\end{equation}
Hence, we expect  the optimum error probability at expected blocklength $N$ to behave as $\exp( - (B/C) N\rho_N)$. Note that the ``exponent'' here contains $\rho_N$ instead of $\rho_n^2$ (for fixed-length codes as discussed in Section~\ref{sec:related}) so  this is further evidence that variable-length codes  dramatically improve the error probability performance over fixed-length codes. This phenomenon has also been observed in other contexts such as decoding with the erasure option~\cite[Theorems~1~\&~3]{HayTan15} and streaming communications with variable decoding delay~\cite[Theorem~7]{LeeTanKhisti16}.
This derivation in \eqref{eqn:heur} is, of course, heuristic and non-rigorous. This paper aims to make  this derivation precise. The contributions are twofold. 
\begin{enumerate}
\item Our first contribution, the direct part, is to judiciously modify   Burnashev's original coding scheme~\cite{Burnashev1976} to achieve~\eqref{eqn:burn} so that it is amenable to analysis in the moderate deviations regime. In particular, we 
derive some new results (e.g., Lemmas~\ref{lemmanew1} and~\ref{lemmanew2}) for  the stopping times of sequences of random variables with properties that resemble both supermartingales and submartingales.
These extensions play   important roles to bound the expectations of the stopping times of the  codes and thus obtaining the exact moderate deviation constant.
\item Our second contribution, the converse part, consists in supplementing some  new real analytical arguments to Burnashev's converse proof in~\cite{Burnashev1976}.  Compared to the original argument~\cite{Burnashev1976}, we also simplify the proof technique, which involves the construction of an appropriate submartingale (cf.\ Lemma~\ref{lem5new}). We do this   by   leveraging   ideas from Burnashev's sequential hypothesis testing paper~\cite{Burnashev80}.
\end{enumerate}
\subsection{Notational Conventions} We use asymptotic notation such as $O(\cdot)$ in the standard manner, e.g., $f_n=O(g_n)$ holds if  $\limsup_{n\to\infty} |f_n/g_n|<\infty$. We use $\ln x$ to denote the natural logarithm so information units throughout are in nats.  We also define the function $(x)_a=x \bone\{x\geq a\}$ for  $x, a\in \bbR$.  The minimum of two  real numbers $a$ and $b$ is denoted interchangeably as $\min\{a,b\}$ and $a\wedge b$. As is usual in information theory, $Z_i^j$ denotes the random vector $(Z_i,Z_{i+1},\ldots, Z_j)$. We usually write $Z_1^j$ as $Z^j$ for brevity. 

For any discrete   sample space $\calZ$, a $\sigma$-algebra $\calF$ on $\calZ$, a random variable  $Z$, and a regular conditional probability measure  $\bbP(\cdot|\calF)$ on $\calZ$, define   the random and usual conditional entropies as 
\begin{align}
\label{eq1notation}
 \calH(Z|\calF)&=-\sum_{z\in \calZ} \bbP (z|\calF)\ln \bbP(z|\calF),\quad\mbox{and}  \\
 H(Z) &=\calH(Z|\sigma(\emptyset,\calZ)). 
\end{align}
If $\calF=\sigma(Y^n)$ for some vector $Y^n$, we write $\sigma(Y^n)$ as $Y^n$ in  \eqref{eq1notation} for simplicity~\cite{Billingsley}.

\subsection{Organization of the Paper}
The rest of this paper is structured as follows. In Section \ref{sec:dmc_bc}, we provide a precise problem statement for DMCs with variable-length codes with  feedback and we state the main result. The achievability proof is provided in Section~\ref{sec:achproof}, and the converse proof is provided in Section~\ref{sec:conveseproof}. Technical derivations are relegated to the appendices.
 
\section{Problem Definition and Main Result}\label{sec:dmc_bc}
\begin{definition}
\label{def1}
A $(M,N)$-{\em variable-length feedback (VLF) code} for a DMC $P_{Y|X}$, where $N$ is a positive real and $M$ is a positive integer, is defined by
\begin{itemize}
\item A set of equiprobable messages $\calW=\{1,2,\ldots,M\}$.
\item A sequence of encoders $\{f_n: \calW \times \calY^{n-1} \to \calX\}_{ n\geq 1}$, defining channel inputs
$
X_n=f_n(W,Y^{n-1})$.
\item A sequence of decoders $\{g_n: \calY^n \to \calW \}_{n\ge 1}$, providing   estimates of $W$ at various times $n$ at the decoder.
\item A   random variable $\tau$, which is  a stopping time of the filtration $\{\sigma(Y^n)\}_{n=0}^{\infty} $. Furthermore, $\tau$ satisfies  
\begin{align}
\label{eq8def}
\bbE(\tau) \leq N.
\end{align}
\end{itemize}
\end{definition}
The final decision at the decoder is computed at time $\tau$ as follows: $\hat{W}=g_{\tau}(Y^{\tau})$. 
The {\em  average error probability} of a given $(M_N,N)$-VLF code with rate $R_N = \frac{1}{N}\ln M_N$ is defined as
$\rvP_{\rme}(R_N,N)=\bbP(\hat{W} \neq W )$.
\begin{definition} \label{def:rel}
The number $E\geq 0$ is an {\em achievable moderate deviations constant} if there exists a family (indexed by $N\in\bbR_+$) of $(M_N,N)$-VLF codes  with $M_N=\exp(NR_N)$  and a   family of positive real numbers $\{\rho_N\}_{N\in \bbR_+}$  satisfying $R_N  \geq C-\rho_N$ and 
\begin{align}
\label{mainreq1}
\lim_{N\to \infty} \rho_N &= 0,\\
\label{mainreq2}
\lim_{N \to \infty} \rho_N \sqrt{N}&=\infty,\\
\lim_{N\to \infty} \rvP_{\rme}(R_N,N)&=0,\\
\label{mainreq4}
\liminf_{N\to \infty}\,\,-\frac{\ln \rvP_{\rme}(R_N,N)}{N\rho_N} &\geq E.
\end{align}
We define the {\em (optimal) moderate deviations constant} $E^*$ (which  implicitly depends on $\{\rho_N\}_{N\in \bbR_+}$ and $P_{Y|X}$)   as the supremum of all  achievable moderate deviations constants.
\end{definition}
\begin{definition} \label{def:info_q} For a given DMC $P_{Y|X}$, we define 
\begin{align}
\label{defineB}
B&=\max_{x,x'\in \calX} D(P_{Y|X}(\cdot|x)\,\|\,P_{Y|X}(\cdot|x')),\\
\label{defineBstar}
B^*&=\max_{ \substack{x,x'\in \calX \\ D(P_{Y|X}(\cdot|x)\,\|\,P_{Y|X}(\cdot|x'))=B }}   D(P_{Y|X}(\cdot|x')\,\|\,P_{Y|X}(\cdot|x))\\
C&=\max_{P_X} I(X;Y),\\
T&=\max_{x,x'\in \calX, y \in \calY}\frac{P_{Y|X}(y|x)}{P_{Y|X}(y|x')}, \label{eqn:defT}\\
\label{defineC2}
C_2&=\max_{x,x' \in \calX, y \in \calY} \left|\ln \frac{P_{Y|X}(y|x)}{P_{Y|X}(y|x')}\right|.
\end{align}
\end{definition}
The main result of this paper is stated in the following theorem.
\begin{theorem} \label{thm:main}
Assume that there exists  a capacity-achieving input distribution $P_X^*$ for the DMC $P_{Y|X}$ satisfying $\min_{x\in\calX}P_X^*(x)>0$. Then for $B<\infty$, the following holds
\begin{align}
E^*=\frac{B}{C}.
\end{align} 
\end{theorem}
\begin{remark} {\em  Some remarks are in order.
\begin{itemize}
\item For the case $B< \infty$, all quantities defined in Definition~\ref{def:info_q} are finite.
\item For the case $B=\infty$, Burnashev~\cite[Section 6]{Burnashev1976} proved the existence of an $(M,N)$-VLF code with zero error probability for any $M, N$ by choosing an appropriate pair of input symbols for the hypothesis testing phase of the proposed coding scheme. This fact obviously means that $E^*=\infty$ when $B=\infty$.

\item The conclusion of Theorem \ref{thm:main} remains the same if the requirement on $\rho_N$ in~\eqref{mainreq2} is relaxed to $\rho_N N^t \to\infty$ for any $t\in(0,1)$    (instead of restricting to $t =1/2$ in~\eqref{mainreq2}). For example, we may take  $\rho_N=N^{-s} $  for any $s\in (0,1)$. Theorem~\ref{thm:main}   and~\eqref{mainreq4} then ensure  that  error probability $\rvP_{\rme}(R_N,N)$ is approximately $\exp(-\Theta(N^{1-s} ))$. However, this does  not include all subexponential functions such as decaying polynomials in which $\rvP_{\rme}(R_N,N)\approx N^{-\kappa}$ for some $\kappa>0$. We state~\eqref{mainreq2} as it is for notational simplicity in the proof. 
\end{itemize}}
\end{remark}
The achievability  and converse proofs of Theorem~\ref{thm:main} are provided in Sections~\ref{sec:achproof} and~\ref{sec:conveseproof} respectively.

\section{Achievability Proof} \label{sec:achproof}
We start with four preliminary lemmas before providing the achievability proof of Theorem \ref{thm:main}.
\begin{lemma} \label{lem6}
Let $K_1,K_2,K_3$ be three positive numbers and let the sequence $\{ \xi_n\}_{n=1}^{\infty}$ be a submartingale adapted to the filtration $\{ \calF_n\}_{n=1}^\infty$, \red{and $\xi_0 \in \bbR$ is a constant}. In addition, assume that
\begin{align}
\label{eqbunarshev1}
\bbE(\xi_{n+1}|\calF_n) &\geq \xi_n +K_1, \quad \mbox{if} \quad \xi_n <0,\\
\label{eqbunarshev2}
\bbE(\xi_{n+1}|\calF_n) &\geq \xi_n +K_2, \quad \mbox{if} \quad \xi_n \geq 0,\\
\label{eqbunarshev3}
|\xi_{n+1}-\xi_n| &\leq K_3,
\end{align}
and the stopping time $\tau$ is given by 
\begin{align}
\label{eqburnashevstopping}
\tau=\inf\{n: \xi_n \geq T\}, 
\end{align} for some $T \in \bbR$. Then, we have
\begin{align}
\label{eq17keynote}
\bbE(\tau) &\leq  K_2^{-1} |T|-K_1^{-1}\xi_0 \bone\{\xi_0 <0\}-K_2^{-1}\xi_0 \bone\{\xi_0\geq 0\}+f(K_1,K_2,K_3),
\end{align}
where the function $f$ depends only on $K_1, K_2$, and $K_3$. 
\end{lemma}
\begin{IEEEproof}[Proof of Lemma~\ref{lem6}]
Please see Appendices \ref{lem6:proof_prelims} and~\ref{lem6:proof}.
\end{IEEEproof} 
\begin{remark} \label{rmk}  {\em  Some remarks concerning Lemma \ref{lem6} are in order.
\begin{itemize}
\item This lemma is an extension of~\cite[Lemma 6]{Burnashev1976} and~\cite[Lemma~1]{Burnashev75} to account for a wider range of parameters. It is   proved by  Naghshvar, Javidi, and Wigger for the case $T>0$~\cite[Lemma~8]{Naghshvar15} \red{and without the constraint~\eqref{eqbunarshev3}}. However, we provide an alternative proof that holds for all $T\in\bbR$ which uses a different  construction of a submartingale (compared to \cite{Burnashev75,Naghshvar15}). We also show rigorously  that $\tau$ is  almost surely finite in Lemma~\ref{lem6a}, which is essential for the proof  of Lemma~\ref{lem6} and of Lemmas~\ref{lemmanew1} and~\ref{lemmanew2} to follow.

\item Moreover, the reason why $\xi_{\tau}$ is  a well-defined random variable was not provided in~\cite{Burnashev1976},~\cite{Burnashev75}, and~\cite{Naghshvar15}. We prove this rigorously in Lemma~\ref{welldefined} in Appendix~\ref{lem6:proof} by  showing that (i) $\xi_\tau$ is a measurable function (due in part to the a.s.\ finiteness of $\tau$)  and (ii) $\xi_\tau\in L^1(\bbR)$ (i.e., $\bbE(|\xi_\tau|  )$ exists and is finite).
\item This lemma, together with Lemmas \ref{lemmanew1} and \ref{lemmanew2} to follow,  is important in bounding the expected lengths of the constructed VLF codes.
\end{itemize}}
\end{remark}
\begin{lemma} \label{lemmanew1} Let $K_1,K_2,K_3$ be three positive numbers and let the sequence of random variables $\{\xi_n\}_{n=1}^{\infty}$ be adapted to a filtration $\{\calF_n\}_{n=1}^{\infty}$. In addition, assume that~\eqref{eqbunarshev3} holds and 
\begin{align}
\label{bunarshev1}
\bbE(\xi_{n+1}|\calF_n) &\geq \xi_n +K_1, \quad \mbox{if}\quad n < \tau_0,\\
\label{bunarshev2}
\bbE(\xi_{n+1}|\calF_n) &\leq \xi_n -K_2, \quad \mbox{if} \quad n \geq \tau_0,
\end{align} where
\begin{align}
\tau_0=\inf\{n: \xi_n \geq T_0\},
\end{align} for some $T_0 \in \bbR$.
Define
\begin{align}
\label{taudefine}
\tau=\inf\{n\geq \tau_0: \xi_n \leq T\},
\end{align} for some $T \leq T_0 \in \bbR$. Then, the following  bound holds
\begin{align}
\label{2020important}
\bbE(\tau-\tau_0) \leq \frac{|T_0-T|+ 3K_3}{K_2}.
\end{align}
\end{lemma}
\begin{IEEEproof} [Proof of Lemma~\ref{lemmanew1}]
Please see Appendix~\ref{lemmanew1:proof}.
\end{IEEEproof}
\begin{lemma} \label{lemmanew2} Assume that all the conditions of  Lemma~\ref{lemmanew1} hold, except that~\eqref{bunarshev2} is replaced by
\begin{align}
\label{bunarshev2a}
\bbE(\xi_{n+1}|\calF_n) \geq \xi_n+K_2, \quad \mbox{if} \quad n \geq \tau_0,
\end{align}
and~\eqref{taudefine} is replaced by~\eqref{eqburnashevstopping} for some $T \geq T_0$ (where $T_0\in\bbR$ is mentioned in Lemma~\ref{lemmanew1}). Then,~\eqref{2020important}  also holds.
\end{lemma}
\begin{IEEEproof}[Proof of Lemma~\ref{lemmanew2}]
This is completely parallel to the proof of Lemma~\ref{lemmanew1} (in  Appendix~\ref{lemmanew1:proof}) and hence omitted.
\end{IEEEproof}
\begin{lemma}\label{extralem} Assume that $\{\rho_L'\}_{L \in \bbR_+}$ is a family of positive numbers satisfying 
\begin{align}
\label{2018new1}
\lim_{L \to \infty} \rho'_L&=0,\\ 
\label{2018new2}
\lim_{L \to \infty} \rho'_L\sqrt{L} &=\infty.
\end{align}
Recall  the definitions of $B, B^*,C_2$ from~\eqref{defineB},~\eqref{defineBstar}, and~\eqref{defineC2}. Let $q_1(P_{Y|X})$ be a function that depends on  $P_{Y|X}$ and let
\begin{align}
\label{defp0L}
p_{0,L}&=1-\frac{1}{L},\\
\label{eq145newest2017}
Z_{0,L}&=\ln \left(\frac{p_{0,L}}{1-p_{0,L}}\right),\\
\label{eq35key2017}
\eps_L&= \exp\left\{\frac{B}{p_{0,L}}\Big[-\frac{L\rho'_L}{C}+\Big(\frac{1}{C}-\frac{p_{0,L}}{B}+\frac{3(1-p_{0,L})}{2B^*}\Big) Z_{0,L} +q_1(P_{Y|X})\Big]\right\},\\
\label{eq38keynew}
A_L&=\frac{Z_{0,L}}{2},\\
\label{eq29key2017}
 p_{1,L}&\in \bigg[0 , \frac{\exp(-Z_{0,L})-\eps_L \exp(-C_2)}{\exp(-A_L)-\eps_L \exp(-C_2)}\bigg].
\end{align}
In addition, assume that $\{W_L\}_{L \in \bbR_+} $ is a family of random variables whose expected values satisfy
\begin{align}
\label{eq24key2017}
p_{0,L}(1-p_{1,L})\bbE(W_L)&=-\frac{p_{0,L}\ln \eps_L}{B} +\frac{\ln (\exp(L(C-\rho'_L))-1)}{C} \nonumber\\
&\quad +\left[\frac{1}{C}-\frac{p_{0,L}}{B}+\frac{(1-p_{0,L})}{B^*}\right]Z_{0,L} +\frac{(1-p_{0,L})|A_L|}{B^*}+q_1(P_{Y|X}).
\end{align}
Then, we have
\begin{align}
\liminf_{L \to \infty} -\frac{\ln \eps_L}{L\rho'_L} &\geq \frac{B}{C} \quad\mbox{and}\\
\bbE(W_L) &\leq L+3\sqrt{L}
\end{align}
for $L$ sufficiently large.
\end{lemma}
\begin{IEEEproof}[Proof of Lemma~\ref{extralem}]
Please see Appendix~\ref{extralem:proof}.
\end{IEEEproof}
\begin{proposition}[Achievability of Theorem~\ref{thm:main}] \label{prop1}
Under the conditions of Theorem~\ref{thm:main},
\begin{align}
E^* \geq \frac{B}{C}.
\label{ach:errexp}
\end{align}
\end{proposition}
\begin{remark} \label{rmk4}  {\em Some remarks are in order.
\begin{itemize}
\item \red{In this achievability proof, our main contribution is to provide a proof for some auxiliary results in~\cite{Burnashev1976} to ensure the arguments carry through for  the moderate deviations regime.  More specifically, in the case $B^* \leq C$, Burnashev~\cite[pp.~260]{Burnashev1976} chose the analogue of $A_L$ in \eqref{eq38keynew} to be negative in his coding scheme to achieve the optimal error exponent~\eqref{eqn:burn}. With this choice, the author stated, without proof, a crucial result~\cite[Eqn.~(5.19)]{Burnashev1976} which is then applied to   derive the optimal error exponent. In the moderate deviations regime, $A_L=(1/2)\ln (p_{0,L}/(1-p_{0,L}))>0$  (cf.~\eqref{eq38keynew}) is chosen to achieve the optimal  moderate deviations constant. Hence, one of the main contributions here is to \red{provide a state and prove Lemmas~\ref{lemmanew1} and~\ref{lemmanew2} which are essential for the proof of the optimal moderate deviations constant.} The different choices of other parameters in~\eqref{defp0L}--\eqref{eq29key2017} vis-\`a-vis the error exponent regime~\cite[Eqn.~(5.22)]{Burnashev1976} also affect our analyses.}
\item It appears (at least to the authors) to be more challenging to adapt Yamamoto-Itoh's coding scheme~\cite{YamamotoItoh1979} compared to Burnashev's coding scheme~\cite{Burnashev1976} since the retransmission probability (or expected length of variable-length code) is a fixed function of the error probability in the communication phase. However, in Burnashev's coding scheme, we can easily control the tradeoff between the error   and   retransmission probabilities by tuning the parameter $A_L$~\cite[pp.~20]{Burnashev1976}.
\end{itemize} }
\end{remark}
\begin{IEEEproof}[Proof of Proposition~\ref{prop1}] We use the same coding scheme as Burnashev~\cite{Burnashev1976}  but our definitions of stopping times  are different. Burnashev's coding scheme consists of two variable-length coding phases. For the sake of completeness, we provide a sketch of his proof and emphasize what we change to make the proof work for the moderate deviations regime.

First we define
\begin{align}
\label{eq18key}
Z_j(n)=\ln \frac{\bbP(W=j|Y^n)}{1-\bbP(W=j|Y^n)}, \quad j=1,2,\ldots,M,
\end{align}
and $Z(n)=Z_{\rmm}(n)$,  
where $\rmm \in \{1,2,\ldots,M\}$ is the transmitted message.  In addition, fix a pair $(x_0,x_0') \in \calX^2$ such that
\begin{align}
\label{cond92}
D(P_{Y|X}(\cdot|x_0)\,\|\,P_{Y|X}(\cdot|x_0'))&=B,\\
\label{cond93}
D(P_{Y|X}(\cdot|x_0')\,\|\,P_{Y|X}(\cdot|x_0))&=B^*.
\end{align}

{\bf Case 1: $B^* >C$:}
For this case, \red{without loss of generality we can assume that there exists a capacity-achieving input distribution $\{P_X^{*}(x)\}_{x \in \calX}$ such that $P_X^{*}(x)>0$ for all $x \in \calX$.\footnote{\red{For the first coding phase, Burnashev~\cite{Burnashev1976}  showed that there exists a coding scheme such that $\bbE[Z(n+1)-Z(n)|Y^n] \geq C,n\in\bbN$ for any capacity-input distribution. This constitutes  one of two key results for the purpose of analyzing the average error probability. The other key result states that $\bbE[Z(n+1)-Z(n)|Y^n] \geq B, n\in\bbN$ holds for the  hypothesis testing phase. The assumption that  such a capacity-achieving input distribution (with full support) exists  makes our notation simpler, i.e, we don't need to change $\calX$ to $\calX'$ where $\calX'$ is the set of all $x \in \calX$ such that $P_X^{*}(x)>0$. Note that $C=D(P_{Y|X}(\cdot|x) \|\sum_{x'} P_X^*(x')P_{Y|X}(\cdot|x'))$ holds for all $x\in \calX'$. }}}  Then, we have~\cite[Theorem~4.5.1]{gallagerIT} the equality $C=D(P_{Y|X}(\cdot|x) \|\sum_{x'} P_X^*(x')P_{Y|X}(\cdot|x'))$ for all $x\in\calX$. 
Define the function, the set, and the constant
\begin{align}
\psi(u,v) &:=\sum_{y \in \calY} P_{Y|X}(y|x_0')\ln \frac{P_{Y|X}(y|x_0') (1-v)}{P_{Y|X}(y|x_0)(1-v)+ u (P_{Y|X}(y|x_0)-P_{Y|X}(y|x_0'))},\\
\calS&:=  \bigg\{ (u,v): \frac{1}{2}\le u\le 1, 0\le v\le 1-u, \psi(u,v)=C \bigg\}, \quad\mbox{and}\\
p_0 &:=\min\left\{u: (u,v) \in \calS \enspace \mbox{for some}\enspace v \in [0,1/2]\right\}. \label{eqn:def_p0}
\end{align}

\emph{1) Encoding:}
\begin{itemize}
\item \emph{Phase 1:} Before transmission at the $(n+1)$-instant, all messages $W \in \{1,2,\ldots,M\}$ are randomly partitioned into $|\calX|$ groups $\calM_1,\calM_2,\ldots,\calM_{|\calX|}$. Here, the probability that message $j \in \{1,2,\ldots,M\}$ will be assigned to the group $\calM_x$ is $\alpha_{j,n}^x/\bbP(W=j|Y^n)$, where $\alpha_{j,n}^x$ (the fraction of $\bbP(W = j|Y^n)$ corresponding to $P_X^*(x)$) is defined as
\begin{align}
\label{modifthe}
\alpha_{j,n}^x:=P_X^*(x)\bbP(W=j|Y^n),
\end{align} for each $x\in \calX$ and $j\in \calM$. It follows from~\eqref{modifthe} that $
\sum_{x\in \cal X}\alpha_{j,n}^x =\bbP(W=j|Y^n)$  and  $
\sum_{j=1}^M \alpha_{j,n}^x=P_X^{*}(x)$. 
\red{The stopping time for phase 1 is defined as
\begin{align}
\label{definetau0}
\tau_0^*=\inf\{n: \max_{j\in \calW}\bbP(W=j|Y^n) \geq p_0\}.
\end{align}}
\item \emph{Phase 2:} \red{At the stopping time $\tau_0^*$,} the posterior probability of one of the messages first exceeds $p_0$, i.e. $\bbP(W={j_0}|Y^n)\geq p_0$ for some $j_0 \in \calW$. Then we subsequently solve a problem of discriminating between two hypotheses: $\rvH_0=\{j_0$ is the true message\} and $\rvH_1=\{j_0$ is a false message\}. Here, $\rvH_0$ is placed in correspondence with input symbol $x_0$, while $\rvH_1$ is placed in correspondence with another symbol $x_0'$, where the pair $(x_0,x_0')$ is chosen in~\eqref{cond92} and~\eqref{cond93}. 
\end{itemize}
The stopping time $\tau^*$ \red{of the phase 2 (or overall coding scheme)} is defined as following:
\begin{align}
\label{eq17key}
\tau^* =\inf\{n \geq 1: \max_{j \in \calW} Z_j(n) \geq  -\ln\eps_N\}
\end{align} 
for a family of real numbers $\{\eps_N\}_{N \in \bbR_+}$ to be determined later; see~\eqref{eqn:eps_N} to follow.

\emph{2) Decoding:}
The decoding is performed at the stopping time $\tau^*$, and the estimated message is
\begin{align}
\hat{W}=\argmax_{k \in \calW} Z_k(\tau_0^*),
\end{align} \red{where $\tau_0^*$ is defined in~\eqref{definetau0}.}

\emph{3)  Moderate Deviations Analysis:}
The error probability $\rvP_{\rme}(R_N,N)$ of this coding scheme can be shown to be bounded above by $\eps_N$~\cite{Burnashev1976}. Moreover, for $B^* >C$, Burnashev~\cite{Burnashev1976} proved that
\begin{align}
\bbE[Z(n+1)-Z(n)|Y^n] &\geq C, \quad \mbox{if}\quad Z(n)< \ln \frac{p_0}{1-p_0},\\
\bbE[Z(n+1)-Z(n)|Y^n] &= B, \quad \mbox{if}\quad Z(n)\geq \ln \frac{p_0}{1-p_0},\\
|Z(n+1)-Z(n)|&\leq C_2.
\end{align}
Now, define
\begin{align}
\tau_{\rmm} =\inf\{n: Z(n) \geq -\ln \eps_N\}.
\end{align}
 Then, applying Lemma~\ref{lem6} with the identifications $K_1=C<K_2=B $, and $K_3=C_2$ for the submartingale $\{\xi_n\}_{n=0}^{\infty} = \{ Z(n)-\ln\frac{p_0}{1-p_0}\}_{n=0}^{\infty}$ adapted to the filtration $\{\calF_n\}_{n=0}^\infty= \{\sigma(Y^n )\}_{n=0}^{\infty}$ with stopping time $\tau_{\rmm}$ and $T=-\ln \eps_N -\ln\frac{p_0}{1-p_0}$, we obtain 
\begin{align}
\bbE(\tau_{\rmm}) 
\label{com1}
&\le B^{-1} \left|-\ln \eps_N -\ln \frac{p_0}{ 1-p_0 }\right|+ C^{-1}  \left|Z(0)-\ln \frac{p_0}{ 1-p_0 } \right|  + h(C,B,C_2),
\end{align}
where    $h(C,B,C_2)$ is a function of $P_{Y|X}$. 
Note that the bound in \eqref{com1} holds because 
\begin{align}
Z(0) =\ln \frac{1/M}{1-1/M}  =-\ln (M-1)\le 0 \leq \ln\frac{p_0}{1-p_0}, \label{com2}
\end{align}
and
$p_0\ge 1/2$ from \eqref{eqn:def_p0} (so $\ln\frac{p_0}{1-p_0}\ge 0$).  

 Now, define new stopping times
\begin{align}
t_j=\inf\{n: Z_j(n) \geq -\ln \eps_N\}, \quad \forall j\in \calW.
\end{align}
Then,  clearly we have
\begin{align}
\label{eq53newnote}
\tau^*=\min_{j \in \calW} t_j.
\end{align}
Now,  choose the family of positive numbers $\{\eps_N\}_{N \in \bbR_+}$ such that
\begin{align}
\label{eq20}
-C^{-1} N\rho_N  -B^{-1}\ln \eps_N+ q_2(P_{Y|X})=0,
\end{align}
where  \red{$q_2(P_{Y|X})$ is defined as
\begin{align}
\label{eq:defq2}
q_2(P_{Y|X}):=C^{-1}\ln \frac{p_0}{1-p_0}-B^{-1}\ln \frac{p_0}{1-p_0}+h(C,B,C_2),
\end{align}
which is a function of the DMC $P_{Y|X}$. We note that since $\eps_N\to 0$, the absolute value in the first term on the right-hand-side of~\eqref{com1} can be removed and that $\ln\frac{p_0}{1-p_0}$ is a constant.} Note that~\eqref{eq20} is equivalent to choosing 
\begin{align}
\eps_N=\exp\left[-\frac{B}{C}N\rho_N +B q_2(P_{Y|X})\right], \quad \forall\, N \in \bbR_+.\label{eqn:eps_N}
\end{align} 
With these choices, we have
\begin{align}
\label{eq:eq56}
\lim_{N\to \infty} \rvP_{\rme}(R_N,N) \leq \lim_{N\to \infty} \eps_N =0.
\end{align}
Therefore, we conclude that
\begin{align}
\label{eqach1}
\liminf_{N \to \infty} \frac{-\ln \rvP_{\rme}(R_N,N)}{N\rho_N} \geq \liminf_{N \to \infty} \frac{-\ln \eps_N}{N\rho_N} = \frac{B}{C} .
\end{align} 
We obtain for $N$ sufficiently large that stopping time $\tau^*$ satisfies
\begin{align}
\bbE(\tau^*)&=\bbE\big(\min_{j\in \calW} t_j\big) \label{eqn:tau_star}\\
\label{keyequation}
&\leq \bbE(\tau_{\rmm})\\
\label{tauarch}
&\leq C^{-1} \ln M -B^{-1}\ln \eps_N+ q_2(P_{Y|X})\\
&=N-C^{-1} N\rho_N  -B^{-1}\ln \eps_N+ q_2(P_{Y|X}) \label{tauarch_logM} \\
&=N,\label{eqn:equalsN}
\end{align}  
where  \eqref{eqn:tau_star} follows from \eqref{eq53newnote},  \eqref{tauarch} follows from~\eqref{com1},~\eqref{com2} and~\eqref{eq:eq56}  \red{with $q_2(P_{Y|X})$ to be defined in~\eqref{eq:defq2}}, \eqref{tauarch_logM} follows from $\log M = N R_N= N(C-\rho_N)$, and \eqref{eqn:equalsN} follows from \eqref{eq20}. 
Thus, the constraint on the expectation of the stopping time is satisfied. This and~\eqref{eqach1} complete  the proof for the case $B^*>C$.


{\bf Case 2: $B^* \leq C$:} 
For this case, we first show that there exists an $(\exp(L(C-\rho'_L)),L+3\sqrt{L})$-VLF code with the average error probability $\rvP'_{\rme}(R_L,L)$ satisfying
\begin{align}
\liminf_{L \to \infty}\frac{-\ln \rvP'_{\rme}(R_L,L)}{L \rho'_L} \geq \frac{B}{C} \label{eqn:first_show}
\end{align}
for any family of  positive     numbers $\{\rho'_L\}_{L \in \bbR_+}$ satisfying~\eqref{2018new1} and~\eqref{2018new2}.  Then we augment the proof to show that $E^*\ge B/C$.

To show \eqref{eqn:first_show}, we modify some choices of important parameters in Burnashev's coding scheme~\cite{Burnashev1976}. These modifications require us to extend some key mathematical results in Burnashev's paper to more general settings (cf.\ Lemmas~\ref{lem6}, \ref{lemmanew1}, and~\ref{lemmanew2}) so they can be applied to the moderate deviations regime. 

As in Burnashev~\cite{Burnashev1976}, for the case $B^*\leq C$, we define
a two-phase encoding scheme is as follows.

\emph{1)  Encoding:}
\begin{itemize}
\item  \emph{Phase 1:} If prior to   instant $n$, the posterior probabilities of all messages $W \in \{1,2,\ldots,\exp(L(C-\rho'_L))\}$ are less than $p_{0,L} \in  [1/2,1)$ defined in~\eqref{defp0L}, the transmission method at the time $n+1$ is the same as the case $B^*> C$ in Case 1 above. \red{The stopping time for phase 1 is defined as
\begin{align}
\label{case2:tau0}
\tau_0^*=\inf\{n: \max_{j\in \calW}\bbP(W=j|Y^n) \geq p_{0,L}\},
\end{align}
where $p_{0,L}\ge 1/2$ is defined in \eqref{defp0L}.}
\item \emph{Phase 2:} \red{At the stopping time $\tau_0^*$}, the posterior probability of one of the messages, indexed by $j_0 \in \calW$, first exceeds $p_{0,L}$, i.e. $\bbP(W=j_0|Y^n)\geq p_{0,L}$   where $p_{0,L}\ge 1/2$ is defined in \eqref{defp0L}. 
 Then we subsequently solve a problem of discriminating between two hypotheses: $\rvH_0=\{j_0$ is the true message\} and $\rvH_1=\{j_0$ is a false message\}. Here, $\rvH_0$ is placed in correspondence with input symbol $x_0$, while $\rvH_1$ is placed in correspondence with another symbol $x_0'$, where the pair $(x_0,x_0')$ is chosen in~\eqref{cond92} and~\eqref{cond93}. 
\end{itemize}

The stopping time \red{of the phase 2 (or the overall coding scheme)} is defined as follows:
\begin{align}
\label{definetau}
\tau^*=\tau_1^*\wedge \tau_2^*,
\end{align} where
\begin{align}
\tau_1^*&=\inf\{n\geq \tau_0^*: \max_{j\in \calW} Z_j \geq  -\ln \eps_L\}, \label{eqn:tau1_star}\\
\tau_2^*&=\inf\{n\geq \tau_0^*: \min_{j \in \calW} Z_j \leq  A_L\},\label{eqn:tau2_star}
\end{align}
where $Z_{0,L}$ and $A_L$ are defined in~\eqref{eq145newest2017} and~\eqref{eq38keynew} respectively. \red{It is easy to see from the definition of $\tau_0^*$ in~\eqref{case2:tau0} and the definition of $Z_{0,L}$ in~\eqref{eq145newest2017} that }
\begin{align}
\label{taustarcase2}
\tau_0^*=\inf\{n: \max_{j \in \calW} Z_j(n) \geq Z_{0,L}\}.
\end{align}
If $\tau_1^*>\tau_2^*$, we retransmit the message. 

\emph{2)  Decoding:}
If $\tau_1^*\leq \tau_2^*$, so $\tau^*=\tau_1^*$, the message is decoded at time $\tau^*$ and the decoding is performed as follows
\begin{align}
\hat{W}=\argmax_{k \in \calW} Z_k(\tau_0^*),
\end{align} where $\tau_0^*$ is defined in~\eqref{taustarcase2}.

\emph{3)  Moderate Deviations Analysis:}
The error probability of this coding scheme $\rvP'_{\rme}(R_L,L)$ can be shown to be bounded above by $\eps_L$~\cite{Burnashev1976}. Assume that $\rmm$ is the transmitted message. Using this coding scheme, Burnashev~\cite{Burnashev1976}  also showed that
\begin{align}
\label{eq112note}
\bbE[Z(n+1)-Z(n)|Y^n] &\geq C,\\
\label{eq112noteb}
|Z_j(n+1)-Z_j(n)|&\leq C_2, \quad \,\,\,\,\forall j \in \calW,\\
\bbE[Z(n+1)-Z(n)|Y^n]&=B,\qquad \,\mbox{if}\quad n \geq \tau_0^*,  \quad \mbox{under} \quad \rvH_0,\\
\label{hypothesish1}
\bbE[Z_{j_0}(n+1)-Z_{j_0}(n)|Y^n]&= -B^*,  \quad \mbox{if} \quad n \geq \tau_0^*, \quad \mbox{under} \quad \rvH_1.
\end{align}
Let $\{Z_{0,L}\}_{L\in\bbR^+}$ and $\{A_L\}_{L \in \bbR^+}$  be  the   families of real numbers defined as in~\eqref{eq145newest2017}   and~\eqref{eq38keynew} respectively. Define
\begin{align}
\label{eq117note}
\tau_{0\rmm}&=\inf\{n \geq 0: Z(n) \geq Z_{0,L}\},\\
\tau_{1\rmm}&=\inf\{n \geq \tau_{0\rmm}: Z(n) \geq -\ln\eps_L \},\\
\tau_{2\rmm}&=\inf\{n \geq \tau_{0\rmm}: Z(n) \leq A_L\},\\
\tau^{-}_{0\rmm}&=\inf\{n \geq 0: \max_{j \in \calW\setminus \{\rmm\}} Z_j(n) \geq Z_{0,L}\},\\
\tau^{-}_{1\rmm}&=\inf\{n \geq \tau^{-}_{0\rmm}: \max_{j \in \calW\setminus \{\rmm\}} Z_j(n) \geq -\ln \eps_L \},\\
\label{definitiontau2m}
\tau^{-}_{2\rmm}&=\inf\{n \geq \tau^{-}_{0\rmm}: \min_{j \in \calW\setminus \{\rmm\}} Z_j(n) \leq A_L\},\\
\tau_{0j_0}&=\inf\{n \geq 0: Z_{j_0}(n) \geq Z_{0,L}\},\\
\label{definitiontau2j_0}
\tau_{2j_0}&=\inf\{n \geq \tau_{0j_0}: Z_{j_0}(n) \leq A_L\}.
\end{align}

Let $p_{1,L}=\bbP(\tau_{2\rmm}< \tau_{1\rmm})$. Using this coding scheme, the retransmission probability\footnote{In the coding scheme, the message will be retransmitted under the condition $\tau_1^*>\tau_2^*$ as stated after \eqref{taustarcase2}.} can be easily obtained as
\begin{align}
\rvP_{\rmx}=1-p_{0,L}(1-p_{1,L}).
\end{align}
It follows that the expected value of the stopping time of the overall coding scheme $\bbE(\tau^*)$ satisfies
\begin{align}
(1-\rvP_{\rmx})\bbE(\tau^*)&= p_{0,L} \bbE(\tau^*|\rvH_0) +(1-p_{0,L})\bbE(\tau^*|\rvH_1)\\
\label{eq88notelan}
&\leq p_{0,L} \bbE(\tau_1^*|\rvH_0) +(1-p_{0,L})\bbE(\tau_2^*|\rvH_1)\\
&= p_{0,L}\bbE(\tau_0^*|\rvH_0)+(1-p_{0,L})\bbE(\tau_0^*|\rvH_1)+ p_{0,L}\bbE(\tau_1^*-\tau_0^*|\rvH_0)+(1-p_{0,L})\bbE(\tau_2^*-\tau_0^*|\rvH_1)\\
\label{eq100notelan}
&= p_{0,L}\bbE(\tau_{0\rmm}|\rvH_0)+(1-p_{0,L})\bbE(\tau^{-}_{0\rmm}|\rvH_1)+ p_{0,L}\bbE(\tau_1^*-\tau_{0\rmm}|\rvH_0)+(1-p_{0,L})\bbE(\tau_2^*-\tau^{-}_{0\rmm}|\rvH_1)\\
\label{eq101notelan}
&\leq p_{0,L}\bbE(\tau_{0\rmm}|\rvH_0)+(1-p_{0,L})\bbE(\tau^{-}_{0\rmm}|\rvH_1)+ p_{0,L}\bbE(\tau_{1\rmm}-\tau_{0\rmm}|\rvH_0)+(1-p_{0,L})\bbE(\tau^{-}_{2\rmm}-\tau^{-}_{0\rmm}|\rvH_1)\\
\label{note126ps}
&\leq p_{0,L}\bbE(\tau_{0\rmm}|\rvH_0)+(1-p_{0,L})\bbE(\tau_{0\rmm}|\rvH_1)+ p_{0,L} \bbE(\tau_{1\rmm}-\tau_{0\rmm}|\rvH_0)+(1-p_{0,L})\bbE(\tau^{-}_{2\rmm}-\tau^{-}_{0\rmm}|\rvH_1)\\
\label{note126}
&= \bbE(\tau_{0\rmm})+ p_{0,L} \bbE(\tau_{1\rmm}-\tau_{0\rmm}|\rvH_0)+(1-p_{0,L})\bbE(\tau^{-}_{2\rmm}-\tau^{-}_{0\rmm}|\rvH_1)\\
\label{eqencod1}
&= \bbE(\tau_{0\rmm})+ p_{0,L} \bbE(\tau_{1\rmm}-\tau_{0\rmm}|\rvH_0)+(1-p_{0,L})\bbE(\tau^{-}_{2\rmm}-\tau_{0j_0}|\rvH_1)\\
\label{eqencod2}
&\leq \bbE(\tau_{0\rmm})+ p_{0,L} \bbE(\tau_{1\rmm}-\tau_{0\rmm}|\rvH_0)+(1-p_{0,L})\bbE(\tau_{2j_0}-\tau_{0j_0}|\rvH_1).
\end{align}
Here,~\eqref{eq88notelan} follows from~\eqref{definetau},~\eqref{eq100notelan} follows from the fact that under $\rvH_0$ we have $\tau_0^*=\tau_{0\rmm}$ and that under $\rvH_1$ we have $\tau_0^*=\tau^{-}_{0\rmm}$,~\eqref{eq101notelan} is obtained from the same arguments as in~\eqref{keyequation},~\eqref{note126ps} follows from the fact that under $\rvH_1$ we have $\tau_{0\rmm} \geq \tau^{-}_{0\rmm}$,~\eqref{eqencod1} follows from the encoding assumption for Phase 2 that $\tau_{0j_0}=\tau^{-}_{0\rmm}$ under $\rvH_1$, and finally~\eqref{eqencod2} follows from the fact that $\tau_{2j_0} \geq \tau^{-}_{2\rmm}$ by using the same arguments as in~\eqref{keyequation}. 

Therefore, we obtain
\begin{align}
\label{note126new}
p_{0,L}(1-p_{1,L})\bbE(\tau^*)&\leq \bbE(\tau_{0\rmm})  + p_{0,L} \bbE(\tau_{1\rmm}-\tau_{0\rmm}) +(1-p_{0,L}) \bbE(\tau_{2j_0}-\tau_{0j_0}).
\end{align}

From~\eqref{eq112note} and~\eqref{eq117note} we see that in Phase 1, $\{ Z(n )\}_{n=0}^{\infty} $ forms a submartingale adapted to the filtration $\{\sigma(Y^n)\}_{n=0}^\infty$ with initial value $Z(0)$ and with stopping time $\tau_{0\rmm}=\inf\{n: Z(n) \geq Z_{0,L}\}$. Applying Lemma~\ref{lem6} with the identifications $K_1=K_2=C$ and $ K_3=C_2$ we obtain
\begin{align}
\label{note130}
\bbE(\tau_{0\rmm}) \leq \frac{Z_{0,L}-Z(0)}{C}+\tilf(C_2,C).
\end{align}
for a function $\tilf$ that depends  only on $C_2$ and $C$. 

In Phase 2, there are two hypotheses.
\begin{itemize}
\item Under $\rvH_1$, the sequence 
\begin{align}
\tilde{Z}_{j_0}(n)=\begin{cases} Z(n),& \tau < \tau_{0\rmm}\\ Z_{j_0}(n),&\tau \geq \tau_{0\rmm}\end{cases},
\end{align} which is adapted to the filtration $\{\sigma(Y^n)\}_{n=0}^{\infty}$, satisfies all requirements of Lemma~\ref{lemmanew1} with $\tau_0=\tau_{0j_0}, \tau=\tau_{2j_0},K_1=C, K_2=B^*, K_3=C_2, T=A_L$, and  $T_0= Z_{0,L}$. Therefore we obtain
\begin{align}
\label{note132b}
\bbE(\tau_{2j_0}-\tau_{0j_0}|\rvH_1) \leq \frac{Z_{0,L}+|A_L|+3C_2}{B^*}.
\end{align}
Note that for $A_L>T$,~\eqref{note132b} still holds since $\tau_{2j_0}=\tau_{0j_0}$. 
\item Under $\rvH_0$, the sequence $\{\tilde{Z}_{j_0}(n)\}_{n=0}^{\infty}$, which is adapted to the filtration $\{\sigma(Y^n)\}_{n=0}^{\infty}$, satisfies all requirements of Lemma~\ref{lemmanew2} with $\tau_0=\tau_{0\rmm}, \tau=\tau_{1\rmm},K_1=C, K_2=B, K_3=C_2, T=-\ln \eps_N$, and  $T_0= Z_{0,L}$. Therefore we obtain
\begin{align}
\label{note132a}
\bbE(\tau_{1\rmm}-\tau_{0\rmm}|\rvH_0) \leq \frac{-\ln \eps_N -Z_{0,L}+3C_2}{B}.
\end{align}
\end{itemize}
Note that
\begin{align}
Z(0)=\ln\left(\frac{1/\exp(L(C-\rho'_L))}{1-1/\exp(L(C-\rho'_L)} \right)=-\ln (\exp(L(C-\rho'_L))-1). \label{notenew}
\end{align}
We choose
\begin{align}
\label{defrhocoma}
\rho'_L=\rho_{L+3\sqrt{L}}-\frac{3}{\sqrt{L}}(C-\rho_{L+3\sqrt{L}}).
\end{align}
It is easy to see that~\eqref{2018new1} and~\eqref{2018new2} hold for $\rho_L'$. In addition, from~\eqref{note126new}--\eqref{notenew}, the expectation of stopping time $\tau^*$ satisfies  
\begin{align}
p_{0,L}(1-p_{1,L})\bbE(\tau^*)&=-p_{0,L}\frac{\ln \eps_L}{B}-\frac{Z(0)}{C}+\left[\frac{1}{C}-\frac{p_{0,L}}{B}+\frac{(1-p_{0,L})}{B^*}\right]Z_{0,L} +\frac{(1-p_{0,L})|A_L|}{B^*}+q_3(P_{Y|X})\\
\label{eq24key2018}
&=-p_{0,L}\frac{\ln \eps_L}{B}+\frac{\ln (\exp(L(C-\rho'_L))-1)}{C}\nonumber\\
&\quad +\left[\frac{1}{C}-\frac{p_{0,L}}{B} +\frac{(1-p_{0,L})}{B^*}\right]Z_{0,L} +\frac{(1-p_{0,L})|A_L|}{B^*}+q_3(P_{Y|X}),
\end{align}
where $q_3(P_{Y|X})$ is depends on  $P_{Y|X}$  and $p_{1,L}=\bbP(\tau_{2\rmm}< \tau_{1\rmm})$ satisfies~\eqref{eq29key2017}. A special case of~\eqref{eq24key2018} was stated without proof in~\cite[Eqn.~(5.19)]{Burnashev1976}, where the author assumed that $A_L<0$~\cite[pp.~260]{Burnashev1976}. The equality~\eqref{eq24key2018} is more general as it holds for all $A_L$; hence we can set $A_L=Z_{0,L}/2>0$ (cf.~\eqref{eq38keynew}). Note that we can easily handle the case $A_L>0$ because of our newly-developed Lemmas~\ref{lemmanew1} and~\ref{lemmanew2} leading  to \eqref{note132b} and \eqref{note132a} respectively.  

 By setting $p_{0,L}$ and $\eps_L$ as~\eqref{defp0L} and~\eqref{eq35key2017} and applying Lemma~\ref{extralem}, we can find an $(\exp(L(C-\rho'_L)), L+3\sqrt{L})$-VLF code such that the average error probability $\rvP'_{\rme}(R_L,L)$ satisfies 
\begin{align}
\liminf_{L\to \infty} \frac{-\ln \rvP'_{\rme}(R_L,L)}{L\rho'_L}\geq \liminf_{L\to \infty} \frac{-\ln \eps_L}{L\rho'_L} \geq \frac{B}{C}.
\end{align}
Thus, \eqref{eqn:first_show} is shown. 
Observe from~\eqref{defrhocoma} that
$
\exp(L(C-\rho'_L))=\exp((L+3\sqrt{L})(C-\rho_{L+3\sqrt{L}})).
$ 
Therefore, there exists an $(\exp((L+3\sqrt{L})(C-\rho_{L+3\sqrt{L}})), L+3\sqrt{L})$-VLF code\footnote{Note that this code has the same average error probability as the $(\exp(L(C-\rho'_L)), L+3\sqrt{L})$-VLF code, i.e., $\rvP_{\rme}(R_{L+3\sqrt{L}},L+3\sqrt{L}) = \rvP'_{\rme}(R_L,L)$. } such that
\begin{align}
\label{eq219newest2017}
\liminf_{L\to \infty} \frac{-\ln \rvP_{\rme}(R_{L+3\sqrt{L}},L+3\sqrt{L})}{(L+3\sqrt{L})\rho_{L+3\sqrt{L}}}&=\liminf_{L\to \infty} \frac{-\ln \rvP_{\rme}(R_{L+3\sqrt{L}},L+3\sqrt{L})}{L\rho'_L}\\
\label{eq220newest2017}
&=\liminf_{L\to \infty} \frac{-\ln \rvP'_{\rme}(R_L,L)}{L\rho'_L}\geq \frac{B}{C}.
\end{align}

By choosing $L$ such that $L+3\sqrt{L}=N$, this is equivalent to assertion of the existence of an $(\exp(N(C-\rho_N)),N)$-VLF code such that 
\begin{align}
\label{eq81lan2017}
\liminf_{N\to \infty} \frac{-\ln \rvP_{\rme}(R_N,N)}{N\rho_N} \geq \frac{B}{C}.
\end{align}
This concludes the proof for the achievability part of Theorem~\ref{thm:main}.
\end{IEEEproof}
\section{Converse Proof}\label{sec:conveseproof}
To prove the converse, we use similar proof arguments as in Burnashev's paper~\cite{Burnashev1976} together with Lemmas~\ref{lemimport} and~\ref{lemconverse} below.  We first state some of Burnashev's lemmas that are used in our converse proof. 
\begin{lemma}
\label{lem1newest}
Under the condition that $\bbP(\tau <\infty)=1$, the following  Fano-type  inequality holds
\begin{align}
\label{eqn:fano}
\bbE\left[\calH(W|Y^{\tau})\right] &\leq h(\rvP_{\rme}(R_N,N))+\rvP_{\rme}(R_N,N)\ln(M-1).
\end{align} 
\end{lemma}
\begin{lemma}
\label{lem3new}
 For any $n\geq 0$ the following inequality holds almost surely
\begin{align}
\bbE[\calH(W|Y^n)- \calH(W|Y^{n+1})|Y^n] &\leq C.
\end{align}
\end{lemma}
\begin{lemma}\label{lem5new} For any $n\geq 0$ the following inequality holds almost surely
\begin{align}
\label{eq66new}
\bbE[\ln\calH(W|Y^n)-\ln \calH(W|Y^{n+1})|Y^n] &\leq B.
\end{align} 
\end{lemma}
\begin{IEEEproof}[Proof of Lemma~\ref{lem5new}]
Please see Appendix~\ref{lem5new:proof} in which we provide a self-contained proof by combining ideas in~\cite[Lemma~3]{Burnashev1976} and Burnashev's sequential hypothesis testing paper~\cite[Lemma~3]{Burnashev80}.
\end{IEEEproof}
\begin{remark}\label{rem:conv}
{\em Some remarks concerning Lemma \ref{lem5new} are in order.
\begin{itemize}
\item This lemma is less strict than (i.e., a generalization of)   the original version in~\cite[Lemma 3]{Burnashev1976}. We have removed the boundedness assumption, i.e.,  that $\calH(W|Y^n) \leq \calH^*(P_{Y|X})$ where $\calH^*(P_{Y|X})$ is  some function of the DMC $ P_{Y|X}$ in~\cite[Lemma 3]{Burnashev1976}.
\item Without the boundedness assumption that $\calH(W|Y^n) \leq \calH^*(P_{Y|X})$, the proof of~\cite[Theorem 1]{Burnashev1976} which leads to the error exponent result in~\eqref{eqn:burn} can be simplified thanks to a simpler construction of an appropriate submartingale.
\end{itemize} }
\end{remark}

\begin{lemma} 
\label{lem7newest}
The following inequality holds almost surely
\begin{align}
\bbE\left[\left(\ln \calH(W|Y^n)-\ln \calH(W|Y^{n+1})\right)_\vartheta\,\middle|\,Y^n\right]\leq \varphi(\vartheta) 
\end{align}
where 
$
\varphi(\vartheta) =\left(\ln T\right)_\vartheta$ and $T$ is defined in \eqref{eqn:defT}.
Under the condition $B<\infty$, $\varphi(\vartheta)=0$ for $\vartheta$ sufficiently large.
\end{lemma}

Now, we present two new lemmas that are useful to analyze the moderate deviations regime. Lemma~\ref{lemimport} is important to deal with the case in which $R_N$ is strictly larger than   $C-\rho_N$; this is allowed by Definition~\ref{def:rel}. Lemma~\ref{lemconverse} is used to show an important result that for any $(\exp(N(C-\rho_N)),N)$-VLF code, \eqref{converseeq145} in the proof of the converse holds. 

\begin{lemma} \label{lemimport}
Let $\{x_n\}_{n\in \bbR_+}$ and $\{y_n\}_{n \in \bbR_+}$ be two  families  of non-negative numbers such that
\begin{align}
\label{eq219newest}
\limsup_{k \to \infty} x_{n_k} &\geq \limsup_{k \to \infty} y_{n_k} \geq 0
\end{align} for any pair of subsequences $(\{x_{n_k}\}_{k=1}^{\infty}, \{y_{n_k}\}_{k=1}^{\infty})$ of the original families $(\{x_n\}_{n\in \bbR_+},\{y_n\}_{n\in \bbR_+})$. In addition, assume that
\begin{align}
\label{xnliminf}
\liminf_{n \to \infty}x_n&=0.
\end{align}
Then, we have
\begin{align}
\label{liminfy}
\liminf_{n \to \infty}y_n=0.
\end{align}
\end{lemma}
\begin{IEEEproof}[Proof of Lemma \ref{lemimport}]
Please see  Appendix~\ref{lemimport:proof}.
\end{IEEEproof}
\begin{lemma}\label{lemconverse}
Assume that $\{\phi_N\}_{N \in \bbR_+}$ is a family of real numbers in the open interval $(0,1)$ satisfying
\begin{align}
 &-\frac{\ln \phi_N}{N\rho_N}-\frac{ \ln (NC-N\rho_N-\ln \phi_N)}{N\rho_N} \nonumber \\
\label{eq37key}
&\quad\leq  \frac{B }{C N\rho_N}+\frac{B \phi_N}{\rho_N}-\frac{B\phi_N}{C} +\frac{B}{C}+\frac{O(1)}{N\rho_N},
\end{align} where $O(1)$ is a bounded constant as $N\to \infty$. Here, $\{\rho_N\}_{N=1}^{\infty}$ is a family of real positive numbers satisfying~\eqref{mainreq1} and~\eqref{mainreq2}. Then, we have
\begin{align}
\label{conversekey}
\limsup_{N\to \infty}-\frac{\ln \phi_N}{N} < \infty.
\end{align}
\end{lemma}
\begin{IEEEproof}[Proof of Lemma \ref{lemconverse}]
Please see Appendix~\ref{lemconverse:proof}.
\end{IEEEproof}
We are now ready to present the proof of the converse. 
\begin{proposition}[Converse of Theorem~\ref{thm:main}]  \label{prop2}
Under the conditions of Theorem~\ref{thm:main},
\begin{align}
\label{converseeq}
E^* \leq \frac{B}{C}.
\end{align}
\end{proposition}
The main contribution of the proof of Proposition \ref{prop2} is to supplement new mathematical analyses that are amenable to the moderate deviations setting. In the following proof, all the steps from~\eqref{rev:153} to~\eqref{rev:173} are different from Burnashev's work~\cite{Burnashev1976}.
\begin{remark}  \label{rem:berlin}
\red{\em We remark that Berlin  {\em et al.}~\cite{Berlin2009a} provided a simple and elegant converse proof of \eqref{eqn:burn} by emulating the two-phase achievability proof in   Yamamoto and Itoh's paper~\cite{YamamotoItoh1979}. The approach in~\cite{Berlin2009a} {\em may possibly} be also used to obtain an upper bound for the optimal moderate deviations constant. However, an optimal way to choose the parameter $\delta$ in~\cite[Theorem 1]{Berlin2009a} is not obvious (at least to the authors). This is because of the presence of the multiplicative term, i.e.,  $(1-\delta-\rvP_{\rme}/\delta)$,  right before $(\ln M)/C$  and the additive term $(\ln(\lambda\delta)-\ln 4)/B$ in~\cite[Eqn.~(18)]{Berlin2009a}. In addition, there are some limiting statements  that have to be modified and it is not clear how to do so in a rigorous and optimal manner (to prove a tight upper bound on~$E^*$).} 
\end{remark}
\begin{IEEEproof}[Proof of Proposition~\ref{prop2}] For $R_N \geq C-\rho_N$, we first realize that if there exists an $(\exp(NR_N),N)$-VLF code with an average error probability $\rvP_{\rme}(R_N,N)$, we can find a $(\exp(N(C-\rho_N)),N)$-VLF code with an average error probability $\rvP_{\rme}(C-\rho_N,N)$ satisfying $\rvP_{\rme}(C-\rho_N,N) \leq \rvP_{\rme}(R_N,N)$ by removing $\exp(NR_N)-\exp(N(C-\rho_N))$ messages with the highest conditional error probabilities. This means that $\inf_{N \in \calA} \rvP_{\rme}(R_N,N) \geq \inf_{N \in \calA} \rvP_{\rme}(C-\rho_N,N)$ for any $\calA\subset \bbR_+$. Hence, 
\begin{align}
\label{eqkeysup}
\limsup_{k\to \infty} -\frac{\ln \rvP_{\rme}(C-\rho_{N_k},N_k)}{N_k\rho_{N_k}}\geq \limsup_{k\to \infty} -\frac{\ln \rvP_{\rme}(R_{N_k},N_k)}{N_k\rho_{N_k}},
\end{align} for any increasing subsequence of positive numbers $\{N_k\}_{k=1}^{\infty}$. To show~\eqref{converseeq} we consider two cases:
\begin{itemize}
\item \emph{Case 1:} There exists a family of $(\exp(N(C-\rho_N)),N)$-VLF codes such that
\begin{align}
\label{newpoint}
\liminf_{N\to \infty} -\frac{\ln \rvP_{\rme}(C-\rho_N,N)}{N\rho_N}=0.
\end{align}
\end{itemize}
It follows that~\eqref{converseeq} trivially holds since from~\eqref{eqkeysup},~\eqref{newpoint}, and Lemma~\ref{lemimport} (with $\phi_N=\rvP_{\rme}(R_N,N)$) we must have
\begin{align}
\liminf_{N\to \infty} -\frac{\ln \rvP_{\rme}(R_N,N)}{N\rho_N}=0,
\end{align}
which leads to
\begin{align}
E^* \leq \liminf_{N\to \infty} -\frac{\ln \rvP_{\rme}(R_N,N)}{N\rho_N}=0 \leq \frac{B}{C}.
\end{align}
\begin{itemize}
\item \emph{Case 2:} For all families of $(\exp(N(C-\rho_N)),N)$-VLF codes, the following holds
\begin{align}
\label{verykey}
\alpha=\liminf_{N\to \infty} -\frac{\ln \rvP_{\rme}(C-\rho_N,N)}{N\rho_N}> 0.
\end{align}
\end{itemize}
For this case, we first prove that
\begin{align}
\label{converseinter}
\limsup_{N\to \infty} -\frac{\ln \rvP_{\rme}(C-\rho_N,N)}{N\rho_N} \leq \frac{B}{C}
\end{align} for any $(\exp(N(C-\rho_N)),N)$-VLF codes. Next, we extend the analysis to $(\exp(NR_N),N)$-VLF codes for $R_N \geq C-\rho_N$.

To prove~\eqref{converseinter} for all $(\exp(N(C-\rho_N)),N)$-VLF codes satisfying~\eqref{verykey}, we use the same converse proof techniques as Burnashev~\cite{Burnashev1976} with some augmented arguments to account for the fact that the code is in the moderate deviations regime. Here, a combination of~\cite{Burnashev1976} and~\cite{Burnashev80} makes the proof that the sequence  $\zeta_n$ (to be defined in \eqref{eqn:xi_def}  in the following) is a submartingale  simpler. For completeness, we provide the entire proof for this case.

To begin with, we show the following inequality
\begin{align}
& -\frac{\ln \rvP_{\rme}(C-\rho_N,N)}{N\rho_N}-\frac{ \ln (NC-N\rho_N-\ln \rvP_{\rme}(C-\rho_N,N))}{N\rho_N} \nonumber \\
\label{eq37keyvui}
&\quad\leq \frac{B }{C N\rho_N}+\frac{B \rvP_{\rme}(C-\rho_N,N)}{\rho_N}-\frac{B\rvP_{\rme}(C-\rho_N,N)}{C} +\frac{B}{C}+\frac{O(1)}{N\rho_N}.
\end{align}
Here, $O(1)$ is a bounded constant as $N\to\infty$. 

It is enough to show that \eqref{eq37keyvui} holds for $N<\infty$, i.e., $\bbE(\tau) <\infty$. We also assume that $B< \infty$, otherwise~\eqref{converseinter} obviously holds. Now, as in  Burnashev's arguments~\cite{Burnashev80}, we consider a random sequence
\begin{align}
\zeta_n =\begin{cases} C^{-1}\calH(W|Y^n)+n,& \; \mbox{if}\; \;\;\calH(W|Y^n) \geq A,\\ B^{-1} \ln\calH(W|Y^n)+b+n,&\;\mbox{if} \;\;\; \calH(W|Y^n) \leq A \end{cases}.  \label{eqn:xi_def}
\end{align} where $A$ is the largest positive root of the following equation in $x$:
\begin{align}
\label{eqkey2000}
\frac{x}{C}=\frac{\ln x}{B}+b.
\end{align} For $b$ sufficiently large, we will show that the sequence $\zeta_n$ forms a submartingle with respect to the filtration $\{\sigma(Y^n)\}_{n=0}^{\infty}$. Note that when $b$ sufficiently large, \eqref{eqkey2000} can be shown to have two distinct positive roots $a,A$ and that $A/a$ can be make arbitrarily large by increasing $b$~\cite[pp.~256]{Burnashev1976}.

Indeed, first we suppose that $\calH(W|Y^n) \leq A$. Then, we obtain
\begin{align}
\bbE\left[\zeta_n-\zeta_{n+1}|Y^n\right] &=-1+\bbE\Big[B^{-1} \ln \calH(W|Y^n) + b-(B^{-1}\ln \calH(W|Y^{n+1})+b)\bone\{\calH(W|Y^{n+1}) \leq A\}\nn\\
&\qquad -C^{-1}\calH(W|Y^{n+1})\bone\{\calH(W|Y^{n+1}) > A\}\Big|Y^n\Big]\\
&\le -1+B^{-1}\bbE\left[\ln\calH(W|Y^n)-\ln \calH(W|Y^{n+1})\,\big|\, Y^n\right] \label{eqn:lessA1a}\\
&\le  -1 +B^{-1} \times B =0.\label{eqn:lessA1b}
\end{align} Here, \eqref{eqn:lessA1a} follows from the fact that $x/C \geq (\ln x)/B + b$ for $x \geq A$~\cite[pp.~256]{Burnashev1976} and \eqref{eqn:lessA1b} follows from   Lemma~\ref{lem5new}.

Now, suppose that $\calH(W|Y^n) > A$. Let $a$ be the smaller of the two positive roots of~\eqref{eqkey2000}. Then, for $b$ sufficiently large, 
\begin{align}
&\bbE\left[\zeta_n-\zeta_{n+1}|Y^n\right]\nn\\*
&=-1+C^{-1}\bbE\left[\calH(W|Y^n)-\calH(W|Y^{n+1})|Y^n\right]\nn\\*
&\quad +\bbE\left[(C^{-1}\calH(W|Y^{n+1})-B^{-1}\ln \calH(W|Y^{n+1})-b)\bone\{\calH(W|Y^{n+1}) \leq A\}|Y^n \right]\label{eqn:geA1a}  \\
&\le \bbE\left[(C^{-1}\calH(W|Y^{n+1})-B^{-1}\ln \calH(W|Y^{n+1})-b)\bone\{\calH(W|Y^{n+1}) \leq A\}|Y^n \right] \label{eqn:147new}\\
&=\bbE\left[(C^{-1}\calH(W|Y^{n+1})-B^{-1}\ln \calH(W|Y^{n+1})-b)\bone\{\calH(W|Y^{n+1}) \leq a\}|Y^n \right]\nn\\*
&\quad + \bbE\left[(C^{-1}\calH(W|Y^{n+1})-B^{-1}\ln \calH(W|Y^{n+1})-b)\bone \{a<\calH(W|Y^{n+1}) \leq A\}|Y^n \right]\\
&\le \bbE\left[(C^{-1}\calH(W|Y^{n+1})-B^{-1}\ln \calH(W|Y^{n+1})-b)\bone \{\calH(W|Y^{n+1}) \leq a\}|Y^n \right]  \label{eq192new2017} \\
&\le B^{-1}\bbE\left[(\ln \calH(W|Y^n)-\ln \calH(W|Y^{n+1}))\bone \{\calH(W|Y^{n+1}) \leq a\}|Y^n \right] \label{eq193lan} \\
&\le B^{-1}\bbE\left[(\ln \calH(W|Y^n)-\ln \calH(W|Y^{n+1}))\bone \Big\{\ln \calH(W|Y^n)-\ln \calH(W|Y^{n+1}) >\ln\Big(\frac{A}{a}\Big)\Big\}\Big|Y^n \right]  \label{eqn:geA1b}  \\
&=B^{-1}\bbE\left[(\ln \calH(W|Y^n)-\ln \calH(W|Y^{n+1}))_{\ln \left(\frac{A}{a}  \right)}\Big|Y^n \right]  \label{eqn:geA1e} \\
&\leq B^{-1}\varphi\left(\ln \Big(\frac{A}{a}\Big)\right) \label{eqn:geA1c}\\
&= 0.  \label{eqn:geA1d}
\end{align} In the above chain of inequalities, \eqref{eqn:geA1a} follows from~\eqref{eqn:xi_def},~\eqref{eqn:147new} follows from Lemma~\ref{lem3new},~\eqref{eq192new2017} follows from the fact that 
$
C^{-1}\calH(W|Y^{n+1})\leq B^{-1}\ln \calH(W|Y^{n+1})+b$ if $a<\calH(W|Y^{n+1}) \leq A$, \eqref{eq193lan} follows from the fact that if $\calH(W|Y^{n+1})\leq a$ and $\calH(W|Y^n) > A$ we have $C^{-1}\calH(W|Y^{n+1})-b\leq C^{-1} a -b =B^{-1} \ln a \leq B^{-1} \ln A \leq B^{-1}\ln \calH(W|Y^n)$, \eqref{eqn:geA1b} follows from the assumption that $\calH(W|Y^n) > A$, \eqref{eqn:geA1e} follows from usage of the notation $(x)_a = x\bone\{x\ge a\}$, and \eqref{eqn:geA1c} and \eqref{eqn:geA1d}  follow  from Lemma~\ref{lem7newest} and the fact that $A/a$ can be made arbitrarily large by increasing $b$. Inequalities~\eqref{eqn:geA1d} and~\eqref{eqn:lessA1b} confirm that $\zeta_n$ forms a submartingale with respect to   $\{\sigma(Y^n)\}_{n=0}^{\infty}$. 

Now, since we know that
\begin{align}
\label{eqn:xi_limit}
\zeta_0=\bbE[\zeta_0] \leq \bbE[\zeta_{n\wedge\tau}]\leq \limsup_{n\to \infty} \bbE[\zeta_{n\wedge \tau}],
\end{align}
it follows that for $N$ sufficiently large we have 
\begin{align}
C^{-1} (NC-N\rho_N)&=\zeta_0^{(1)} \label{eqn:ach_rate_a1} \\
&\leq \limsup_{n\to \infty} \bbE[\zeta_{n\wedge \tau}]\label{eqn:ach_rate_a1_b} \\
&  \leq C^{-1}\limsup_{n\to \infty} \bbE\left[\calH(W|Y^{\tau\wedge n})\bone\{\calH(W|Y^{\tau\wedge n})\geq A\}\right]\nn\\
&\quad + \limsup_{n\to \infty} \bbE\left[\tau \wedge n\right]  + \limsup_{n\to \infty} B^{-1} \bbE\left[\ln\calH(W|Y^{\tau\wedge n})\bone\{\calH(W|Y^{\tau\wedge n})\leq A\} \right]+b\label{eqn:ach_rate_a0} \\
&\leq C^{-1}\limsup_{n\to \infty} \bbE\left[\calH(W|Y^{\tau \wedge n})\right]\nn\\
&\quad + \limsup_{n\to \infty} \bbE\left[\tau \wedge n\right]  + \limsup_{n\to \infty} B^{-1} \bbE\left[\ln\calH(W|Y^{\tau\wedge n})\bone\{\calH(W|Y^{\tau\wedge n})\leq A\} \right]+b\\
&\le  C^{-1}\limsup_{n\to \infty} \bbE\left[\calH(W|Y^{\tau\wedge n})\right]  + \limsup_{n\to \infty} \bbE\left[\tau \wedge n\right]  + \limsup_{n\to \infty} B^{-1} \ln \bbE\left[\calH(W|Y^{\tau\wedge n}) \right]+b \label{eqn:ach_rate_a}\\
&=C^{-1}\bbE\left[\calH(W|Y^{\tau})\right]+\bbE\left[\tau\right]  +  B^{-1} \bbE\left[\ln\calH(W|Y^{\tau})\right]\\
&\le  C^{-1}[1+\rvP_{\rme}(C-\rho_N,N)(NC-N\rho_N)] + \bbE\left[\tau\right]+ B^{-1} \ln [h(\rvP_{\rme}(C-\rho_N,N))\nonumber\\
&\quad +\rvP_{\rme}(C-\rho_N,N) (NC-N\rho_N)]+b \label{eqn:ach_rate_b}\\
&= C^{-1}[1+\rvP_{\rme}(C-\rho_N,N)(NC-N\rho_N)] + \bbE\left[\tau \right]+ B^{-1} \ln [-\rvP_{\rme}(C-\rho_N,N)\ln \rvP_{\rme}(C-\rho_N,N) \nn\\*
&\quad -(1-\rvP_{\rme}(C-\rho_N,N))\ln(1-\rvP_{\rme}(C-\rho_N,N))+\rvP_{\rme}(C-\rho_N,N) (NC-N\rho_N)]+b\\
&\le   C^{-1}[1+\rvP_{\rme}(C-\rho_N,N)(NC-N\rho_N)] + \bbE\left[\tau \right]+ B^{-1} \ln [-\rvP_{\rme}(C-\rho_N,N)\ln \rvP_{\rme}(C-\rho_N,N)\nonumber\\
&\quad +\frac{1}{e}+\rvP_{\rme}(C-\rho_N,N) (NC-N\rho_N)]+b\label{eqn:ach_rate_c}\\
&= C^{-1}[1+\rvP_{\rme}(C-\rho_N,N)(NC-N\rho_N)] + \bbE\left[\tau\right]+ B^{-1} \ln [-\rvP_{\rme}(C-\rho_N,N)\ln \rvP_{\rme}(C-\rho_N,N)\nonumber\\
&\quad +\rvP_{\rme}(C-\rho_N,N) (NC-N\rho_N)]+O(1)\label{eqn:ach_rate_d} \\
&\leq C^{-1}[1+\rvP_{\rme}(C-\rho_N,N)(NC-N\rho_N)] + N + B^{-1}\ln \rvP_{\rme}(C-\rho_N,N) \nonumber\\*
&\quad + B^{-1} \ln (NC-N\rho_N-\ln \rvP_{\rme}(C-\rho_N,N))+O(1). \label{eqn:last_rate}
\end{align} Here,  \eqref{eqn:ach_rate_a1} follows from~\eqref{eqn:xi_def} and $\calH(W|Y^{0})=H(W)=NC-N\rho_N$, \eqref{eqn:ach_rate_a1_b} follows from~\eqref{eqn:xi_limit},  \eqref{eqn:ach_rate_a0} follows from~\eqref{eqn:xi_def} and~\eqref{eqn:xi_limit}, \eqref{eqn:ach_rate_a} follows from the fact that for any random variable $U$,  $\bbE[(\ln U) \bone\{U\leq u\}]\leq \ln \bbE(U)$ for all $u \geq 1$ (which is assured by taking $b$ sufficiently large so $A$ eventually becomes larger than $1$),  
   \eqref{eqn:ach_rate_b}  follows from Lemma~\ref{lem1newest} for the   case that $R_N=C-\rho_N$,~\eqref{eqn:ach_rate_c}   follows from the fact that $-x\ln x\leq 1/e$ for $0\leq x\leq 1$,~\eqref{eqn:ach_rate_d}  follows from the fact that $b <\infty$, and~\eqref{eqn:last_rate} follows from~\eqref{eq8def}.

It follows from~\eqref{eqn:last_rate} that
\begin{align}
\label{rev:153}
- B^{-1} \ln \rvP_{\rme}(C-\rho_N,N) &\leq C^{-1}[1+\rvP_{\rme}(C-\rho_N,N) N(C-\rho_N)] +C^{-1}N\rho_N \nonumber\\
&\quad + B^{-1} \ln (N(C-\rho_N)-\ln \rvP_{\rme}(C-\rho_N,N))+O(1).
\end{align}
Hence, we have
\begin{align}
& -\frac{\ln \rvP_{\rme}(C-\rho_N,N)}{N\rho_N}-\frac{ \ln (NC-N\rho_N-\ln \rvP_{\rme}(C-\rho_N,N))}{N\rho_N} \nonumber \\
 &\quad\le \frac{B }{C N\rho_N}+\frac{B \rvP_{\rme}(C-\rho_N,N)}{\rho_N}-\frac{B\rvP_{\rme}(C-\rho_N,N)}{C} +\frac{B}{C}+\frac{O(1)}{N\rho_N},
\end{align}
and~\eqref{eq37keyvui} is shown.

Now, from the assumption in~\eqref{verykey}, we have
\begin{align}
-\frac{\ln \rvP_{\rme}(C-\rho_N,N)}{N \rho_N} \geq \frac{\alpha}{2}
\end{align} for $N$ sufficiently large.  It follows that
\begin{align}
\rvP_{\rme}(C-\rho_N,N) \leq \exp\left(-\frac{\alpha}{2} N\rho_N\right).
\end{align}
Hence, we have
\begin{align}
\label{converseeq3}
0 \leq \limsup_{N\to \infty} \frac{\rvP_{\rme}(C-\rho_N,N) }{\rho_N} \leq \limsup_{N\to \infty}\frac{\exp\left(-\frac{\alpha}{2} N\rho_N\right)}{\rho_N}
\end{align}
Now, we note that for any $\beta>0$, $g_{\beta}(x )=\frac{1}{x}{\exp(-\beta x)}$
   is a decreasing function in $x \in (0,\infty)$. Moreover, since $\sqrt{N}\rho_N \to \infty$ as $N\to \infty$, there exists $N_0$ sufficiently large such that $\sqrt{N} \rho_N \geq 1$ for all $N\geq N_0$. It follows that for all $N\geq N_0$ we have
\begin{align}
\frac{\exp(-\frac{\alpha}{2}\sqrt{N}  (\sqrt{N} \rho_N))}{\sqrt{N}\rho_N} =g_{\frac{\alpha}{2}\sqrt{N}}(\sqrt{N} \rho_N) 
 \leq g_{\frac{\alpha}{2}\sqrt{N}}(1)
=\exp\left(-\frac{\alpha}{2}\sqrt{N}\right).\label{converseeq2}
\end{align}
Hence, we obtain
\begin{align}
0\leq \liminf_{N\to \infty}\frac{\exp\left(-\frac{\alpha}{2} N\rho_N\right)}{\rho_N}\leq \limsup_{N\to \infty}\frac{\exp\left(-\frac{\alpha}{2} N\rho_N\right)}{\rho_N}\leq \limsup_{N\to \infty} \left[\exp\left(-\frac{\alpha}{2}\sqrt{N}\right)\right]\sqrt{N} =0.\label{converseeq4}
\end{align}
From~\eqref{converseeq3} and~\eqref{converseeq4} we obtain
\begin{align}
\label{eq128key}
\lim_{N\to \infty} \frac{\rvP_{\rme}(C-\rho_N,N) }{\rho_N}=\lim_{N\to \infty}\left[\exp\left(-\frac{\alpha}{2}\sqrt{N}\right)\right]\sqrt{N}=0.
\end{align}
By taking lim supremum both sides of~\eqref{eq37keyvui} and using~\eqref{eq128key}, we have
\begin{align}
\label{eq57key}
\frac{B}{C} &\geq \limsup_{N\to \infty}\left[ -\frac{\ln \rvP_{\rme}(C-\rho_N,N)}{N\rho_N}\right]+\liminf_{N\to \infty}\left[-\frac{ \ln (NC-N\rho_N)-\ln \rvP_{\rme}(C-\rho_N,N))}{N\rho_N} \right]\\
\label{converse58key}
&=\limsup_{N\to \infty}\left[ -\frac{\ln \rvP_{\rme}(C-\rho_N,N)}{N\rho_N}\right]-\limsup_{N\to \infty}\left[\frac{ \ln (NC-N\rho_N)-\ln \rvP_{\rme}(C-\rho_N,N))}{N\rho_N} \right].
\end{align}

Next, we will show that
\begin{align}
\label{converseeq145}
\lim_{N\to \infty}\left[\frac{ \ln (NC-N\rho_N)-\ln \rvP_{\rme}(C-\rho_N,N))}{N\rho_N} \right]=0.
\end{align}
and so the second term in  \eqref{converse58key} is zero.   Observe that
\begin{align}
\frac{ \ln (NC-N\rho_N-\ln \rvP_{\rme}(C-\rho_N,N))}{N\rho_N} &=\frac{ \ln (NC-N\rho_N-\ln \rvP_{\rme}(C-\rho_N,N))}{N\rho_N} \\
\label{converseeq147}
&\geq \frac{ \ln\left( C-\rho_N\right)}{\sqrt{N}({\sqrt{N}}\rho_N)}+\frac{\ln N}{\sqrt{N}}\cdot \frac{1}{\sqrt{N}\rho_N} 
\end{align}
Here,~\eqref{converseeq147} follows from the fact that $0< \rvP_{\rme}(C-\rho_N,N)\leq 1$. It follows that
\begin{align}
\liminf_{N\to \infty} \frac{ \ln (NC-N\rho_N-\ln \rvP_{\rme}(C-\rho_N,N))}{N\rho_N}\geq \liminf_{N\to \infty} \frac{ \ln\left(C-\rho_N\right)}{\sqrt{N}({\sqrt{N}}\rho_N)}+\frac{\ln N}{\sqrt{N}} \cdot \frac{1}{\sqrt{N}\rho_N}=0 
\label{liminfzero}
\end{align}
Now, from~\eqref{eq37keyvui} and Lemma~\ref{lemconverse}, we have
\begin{align}
\label{keynewforconverse}
\limsup_{N \to \infty} -\frac{\ln \rvP_{\rme}(C-\rho_N,N)}{N}< \infty.
\end{align}
It follows from~\eqref{keynewforconverse} we have for $N$ sufficiently large that
\begin{align}
-\frac{\ln \rvP_{\rme}(C-\rho_N,N)}{N} \leq \nu,
\end{align} for some constant  $\nu\in (0,+\infty)$. Hence, we have
\begin{align}
\frac{ \ln (NC-N\rho_N-\ln \rvP_{\rme}(C-\rho_N,N))}{N\rho_N} &=\frac{ \ln (NC-N\rho_N-\ln \rvP_{\rme}(C-\rho_N,N))}{N\rho_N} \\
&\leq
\frac{ \ln (C-\rho_N+\nu)}{\sqrt{N} (\sqrt{N}\rho_N)}+\frac{\ln N}{\sqrt{N}}\cdot \frac{1}{\sqrt{N}\rho_N}
\end{align}
It follows that
\begin{align}
\limsup_{N\to \infty} \frac{ \ln (NC-N\rho_N-\ln \rvP_{\rme}(C-\rho_N,N))}{N\rho_N}&\leq \limsup_{N\to \infty} \frac{ \ln (C-\rho_N+\nu)}{\sqrt{N} (\sqrt{N}\rho_N)}+\frac{\ln N}{\sqrt{N}} \cdot \frac{1}{\sqrt{N}\rho_N}\\
\label{limsupzero}
&=0.
\end{align}
Combining~\eqref{liminfzero} and~\eqref{limsupzero}, we obtain~\eqref{converseeq145}. From~\eqref{converse58key} and~\eqref{converseeq145}, we obtain~\eqref{converseinter} as desired.  

Finally, for any $(\exp(NR_N),N)$-VLF codes, by combining~\eqref{mainreq4},~\eqref{eqkeysup}, and~\eqref{converseinter} we have
\begin{align}
\label{rev:173}
E^*  \leq \liminf_{N\to \infty}\left[ -\frac{\ln \rvP_{\rme}(R_N,N)}{N\rho_N}\right]
 \leq \limsup_{N\to \infty}\left[ -\frac{\ln \rvP_{\rme}(R_N,N)}{N\rho_N}\right]
 \leq \limsup_{N\to \infty}\left[ -\frac{\ln \rvP_{\rme}(C-\rho_N,N)}{N\rho_N}\right]
 \leq \frac{B}{C},
\end{align}
concluding the proof for the converse.
\end{IEEEproof}

\numberwithin{equation}{section}
\renewcommand{\theequation}{\thesection.\arabic{equation}}

\appendices
\section{Preliminaries for the Proof of Lemmas \ref{lem6}, \ref{lemmanew1}, and \ref{lemmanew2} }\label{lem6:proof_prelims}
To prove Lemma~\ref{lem6}, \ref{lemmanew1}, and \ref{lemmanew2}, we first state and prove some useful preliminary definitions and lemmas.
\begin{definition}\label{def5}
Let $\{U_n\}_{n=1}^{\infty}$ be a sequence of random variables. The sequence $\{U_n\}_{n=1}^{\infty}$ is called $*$-\emph{submixing} sequence adapted to a filtration $\{\calF_n\}_{n=1}^{\infty}$ if $U_n \in \calF_n$ and that there exists a positive number $N$ and a non-negative function $f$ defined on the integers $n\geq N$ such that $f(n) \to 0$ as $n\to \infty$ and for all $n\geq N, m\geq 1$,
\begin{align}
\label{mixingdef}
|\bbE(U_{n+m}|\calF_m)-\bbE(U_{n+m})|\leq f(n) \bbE|U_{n+m}|.
\end{align}
\end{definition}
%
\begin{lemma}~\cite[Theorem 2.18]{HallHeyde} \label{importantlem} Let $\{W_i\}_{i=1}^{\infty}$ be a sequence of random variables such that
 $\{S_n=\sum_{i=1}^n W_i, \calF_n\}_{n=1}^{\infty}$ is a martingale, and let $\{V_n\}_{n=1}^{\infty}$ be a non-decreasing sequence of positive random variables such that $V_n \in \calF_{n-1}$ for each $n$. Fix $1\leq p\leq 2$. Then
\begin{align}
\label{eq217hey}
\lim_{n \to \infty} V_n^{-1} S_n=0, \quad \mbox{a.s.}
\end{align}
on the event
\begin{align}
\label{usedcondition}
\bigg\{\lim_{n\to \infty} V_n =\infty, \sum_{i=1}^{\infty} V_i^{-p} \bbE[|W_i|^p|\calF_{i-1}]<\infty\bigg\}.
\end{align}
\end{lemma}
\begin{lemma} \label{Heydelema} Let $\{U_n\}_{n=1}^{\infty}$ be a $*$-submixing sequence adapted to a filtration $\{\calF_n\}_{n=1}^{\infty}$ such that $\bbE(U_n)=0$ and $\bbE(U_n^2)< \infty, n\geq 1$. Suppose that
$
\sum_{n=1}^{\infty} b_n^{-2} \bbE(U_n^2) <\infty,
$
and
$
\sup_{n} b_n^{-1} \sum_{i=1}^{n} \bbE|U_i| <\infty,
$
where $\{b_n\}_{n=1}^{\infty}$ is a sequence of positive constants increasing to $\infty$. Then,
\begin{align}
b_n^{-1}\sum_{i=1}^{n} U_i \to 0, \quad \mbox{a.s.}  \label{eqn:lemma12_concl}
\end{align}
\end{lemma}
\begin{remark} {\em Some remarks concerning Lemma \ref{Heydelema} are in order.
\begin{itemize}
\item This lemma is a generalization of~\cite[Theorem 2.20]{HallHeyde}, which considered the case where $\calF_n=\sigma(U_1^n)$ and that $\{U_n\}_{n=1}^{\infty}$ is a $*$-mixing sequence in the sense that for any $B \in \calF_{\infty}=\lim_{n\to \infty} \calF_n$,
\begin{align}
|\bbP(B|\calF_n)-\bbP(B)|\leq f(n) \bbP(B) \quad\mbox{a.s.}
\end{align}
\item By setting $\bone\{B\}=U_{n+m}$, it is easy to see that any $*$-mixing sequence is a $*$-submixing sequence.
\end{itemize}}
\end{remark}
\begin{IEEEproof}[Proof of Lemma~\ref{Heydelema}] \label{Heydelema:proof} The proof is based on~\cite[Theorem 2.20]{HallHeyde}. There are some important changes  to account for the fact that $U_{in_0+j} \notin \sigma \big(\cup_{k=0}^{i}\big(\calF_{kn_0+j}\setminus\calF_{kn_0+j-1}\big) \big)$  for some triplet $(i,j,n_0) \in \bbN^3$. The proof in~\cite[Theorem 2.20]{HallHeyde} is based on the fact that $U_{in_0+j} \in \sigma \big(\cup_{k=0}^{i}\big(\calF_{kn_0+j}\setminus\calF_{kn_0+j-1}\big) \big)$ for any triplet $(i,j,n_0) \in \bbN^3$, which holds when  $\calF_n=\sigma(U_1^n)$.

Given $\eps>0$, there exist an $n_0\geq N$ such that $f(n)<\eps$ for all $n\geq n_0$ since $f(n) \to 0$ as $n\to \infty$.  From~\eqref{mixingdef} we deduce that for all positive integers $i$ and $j$,
\begin{align}
\label{eq1952000}
\Big|\bbE\big[U_{in_0+j}\big|\calF_{(i-1)n_0+j}\big]\Big|&=\Big|\bbE\Big[\bbE \big[U_{in_0+j}\big|\calF_{(i-1)n_0+j}\big] -\bbE\big[U_{in_0+j}\big]\Big|\calF_{(i-1)n_0+j}\Big]\Big|\\
\label{eq1962000}
&\leq \bbE\Big[\big|\bbE\big[U_{in_0+j}\big|\calF_{(i-1)n_0+j}\big] -\bbE\big[U_{in_0+j}\big]\big|\Big|\calF_{(i-1)n_0+j}\Big]\\
\label{eq197new2000}
&\leq \bbE\Big[f(n_0) \bbE|U_{in_0+j}|\Big|\calF_{(i-1)n_0+j}\Big]\\
\label{verynewstar}
&=f(n_0) \bbE|U_{in_0+j}|,
\end{align} almost surely. Here,~\eqref{eq1952000} follows from tower property of conditional expectation~\cite{Billingsley} and the assumption that $\bbE(U_n)=0$ for all $n\in\bbN$, and~\eqref{eq197new2000} follows from~\eqref{mixingdef}.

If $n\geq n_0$, choose nonnegative integers $q$ and $r$ such that $0\leq r\leq n_0-1$ and $n=qn_0+r$. Then, 
\begin{align}
\label{eq1972000}
b_n^{-1} \sum_{i=1}^n U_i &=b_n^{-1}\sum_{i=1}^{n_0} U_i + \sum_{j=0}^{n_0-1} b_n^{-1} \sum_{i=1}^{q-1}  U_{in_0+j}+b_n^{-1} \sum_{j=0}^r U_{qn_0+j}.
\end{align}
Observe that
\begin{align}
&\sum_{j=0}^{n_0-1} b_n^{-1} \sum_{i=1}^{q-1}  U_{in_0+j}+b_n^{-1} \sum_{j=0}^r U_{qn_0+j} \nonumber \\
&=\sum_{j=0}^{n_0-1}b_n^{-1} \sum_{i=1}^{q-1}\Big[U_{in_0+j}-\bbE\big[U_{in_0+j}\big|\calF_{(i-1)n_0+j}\big]\Big] \nonumber \\
&\quad + \sum_{j=0}^r b_n^{-1} \Big[U_{qn_0+j}-\bbE\big[U_{qn_0+j}\big|\calF_{(q-1)n_0+j}\big]\Big] \nonumber \\
&\quad + b_n^{-1} \sum_{j=0}^{n_0-1} \sum_{i=1}^{q-1}  \bbE\big[U_{in_0+j}\big|\calF_{(i-1)n_0+j}\big]+b_n^{-1} \sum_{j=0}^r \bbE\big[U_{qn_0+j}\big|\calF_{(q-1)n_0+j}\big]\\
&\leq \sum_{j=0}^{n_0-1} \sum_{i=1}^{q-1}b_n^{-1}\Big[U_{in_0+j}-\bbE\big[U_{in_0+j}\big|\calF_{(i-1)n_0+j}\big]\Big] \nonumber \\
&\quad + \sum_{j=0}^r b_n^{-1} \Big[U_{qn_0+j}-\bbE\big[U_{qn_0+j}\big|\calF_{(q-1)n_0+j}\big]\Big] \nonumber \\
\label{eq23note}
&\quad + f(n_0) b_n^{-1} \sum_{i=n_0}^n \bbE|U_i|,
\end{align} where~\eqref{eq23note} follows from~\eqref{verynewstar}.

Hence, we obtain
\begin{align}
b_n^{-1} \sum_{i=1}^n U_i &\leq b_n^{-1}\sum_{i=1}^{n_0-1} U_i + \sum_{j=0}^{n_0-1} \sum_{i=1}^{q-1}b_n^{-1}\Big[U_{in_0+j}-\bbE\big[U_{in_0+j}\big|\calF_{(i-1)n_0+j}\big]\Big] \nonumber \\
&\quad + \sum_{j=0}^r b_n^{-1} \Big[U_{qn_0+j}-\bbE\big[U_{qn_0+j}\big|\calF_{(q-1)n_0+j}\big]\Big] \nonumber \\
\label{eq23notenew}
&\quad + f(n_0) b_n^{-1} \sum_{i=n_0}^n \bbE|U_i|.
\end{align}
Since $n_0$ is a fixed  number, it is easy to see that the first term on the right-hand side of~\eqref{eq23notenew} almost surely converges  to zero as $n\to \infty$, i.e.,
\begin{align}
\label{beauty1}
\lim_{n \to \infty} b_n^{-1}\sum_{i=1}^{n_0} U_i=0, \quad \mbox{a.s.}
\end{align}
 Now, for each fixed pair $j \in \{0,1,2,\ldots, n_0-1\}$ and $i \in \bbN$ denote by
\begin{align}
W_{j, i}&=U_{in_0+j}-\bbE\big[U_{in_0+j}\big|\calF_{(i-1)n_0+j}\big],\\
V_{j, i}&=b_{in_0+j}.
\end{align}
Then, we have $V_{j,i} \in \calF_{(i-1)n_0+j}$ for all $i,j$. On the other hand, since $U_n \in \calF_n$ for all $n=0,1,2,\ldots$ we also have
\begin{align}
W_{j,i}=U_{in_0+j}-\bbE\big[U_{in_0+j}\big|\calF_{(i-1)n_0+j}\big] \in \calF_{in_0+j}.
\end{align}
It follows that for each fixed $j \in \{0,1,2,\ldots,n_0-1\}$  we have
\begin{align}
&\bbE\Big[\sum_{i=1}^{q-1} W_{j,i}\Big|\calF_{(q-2)n_0+j}\Big]\nonumber\\
&=\bbE\Big[\sum_{i=1}^{q-2} W_{j,i}\Big|\calF_{(q-2)n_0+j}\Big]
+ \bbE\big[ W_{j,q-1}\big|\calF_{(q-2)n_0+j}\big]\\
&=\sum_{i=1}^{q-2} W_{j,i}+\bbE\big[ W_{j,q-1}\big|\calF_{(q-2)n_0+j}\big]\\
&=\sum_{i=1}^{q-2} W_{j,i}+\bbE\Big[U_{(q-1)n_0+j}-\bbE\big[U_{(q-1)n_0+j}\big|\calF_{(q-2) n_0+j}\big] \Big|\calF_{(q-2)n_0+j}\Big]\\
&=\sum_{i=1}^{q-2} W_{j,i},
\end{align}
or that $\big\{\big(\sum_{i=1}^{q-1} W_{j,i}, \calF_{(q-1)n_0+j} \big)\big\}_{q=1}^{\infty}$ forms a martingale. In addition, we also have
\begin{align}
&\sum_{i=1}^{\infty}  V_{j,i}^{-2} \bbE[W_{j,i}^2|\calF_{(i-1)n_0+j}]\nonumber\\
&=\sum_{i=1}^{\infty}   b_{i n_0+j}^{-2} \bbE\Big[\Big(U_{in_0+j}-\bbE \big[U_{in_0+j}\big|\calF_{(i-1)n_0+j} \big]\Big)^2\Big|\calF_{(i-1)n_0+j}\Big]\\
&\leq \sum_{i=1}^{\infty}   b_{i n_0+j}^{-2} \bbE\big[U_{in_0+j}^2 \big|\calF_{(i-1)n_0+j}\big].
\end{align}
Hence, we obtain that for each fixed $j\in \{0,1,2,\ldots,n_0-1\}$ that
\begin{align}
\bbE\Big[\sum_{i=1}^{\infty}  V_{j,i}^{-2} \bbE[W_{j,i}^2|\calF_{(i-1)n_0+j}]\Big] & = \bbE\Big[\sum_{i=1}^{\infty} b_{i n_0+j}^{-2} \bbE\big[U_{in_0+j}^2\big|\calF_{(i-1)n_0+j}\big]\Big] \\
& =\sum_{i=1}^{\infty}  b_{i n_0+j}^{-2}\bbE[U_{in_0+j}^2] \\
& \leq \sum_{n=1}^{\infty} b_n^{-2} \bbE[U_n^2] \\
&< \infty.
\end{align}
Therefore, it holds almost surely for each fixed $j \in \{0,1,2,\ldots,n_0-1\}$ that
\begin{align}
\sum_{i=1}^{\infty} V_{j,i}^{-2} \bbE[W_{j,i}^2|\calF_{(i-1)n_0+j}]< \infty.
\end{align}
This means that the sequences of random variables $\{W_{j,i}\}_{i=1}^{\infty}$ and $\{V_{j,i}\}_{i=1}^{\infty}$ satisfy~\eqref{usedcondition} in Lemma \ref{importantlem}. It follows that as $q\to \infty$ we have
\begin{align}
 b_{(q-1)n_0+j}^{-1} \sum_{i=1}^{q-1}  \Big[U_{in_0+j}-\bbE\big[U_{in_0+j}\big|\calF_{(i-1)n_0+j}\big]\Big] 
=V_{j,q-1}^{-1}\sum_{i=1}^{q-1} W_{j,i}\to 0, \quad \mbox{a.s.} \label{beauty2}
\end{align} for each fixed $j \in \{0,1,2,\ldots,n_0-1\}$. It follows that as $q \to \infty$
\begin{align}
b_{(q-1)n_0+j}^{-1}\Big| \sum_{i=1}^{q-1}  \Big[U_{in_0+j}-\bbE\big[U_{in_0+j}\big|\calF_{(i-1)n_0+j}\big]\Big]\Big| \to 0, \quad \mbox{a.s.}
\end{align} for each fixed $j \in \{0,1,2,\ldots,n_0-1\}$. Therefore, we obtain
\begin{align}
\sum_{j=0}^{n_0-1} b_{(q-1)n_0+j}^{-1}\Big|\sum_{i=1}^{q-1}  \Big[U_{in_0+j}-\bbE\big[U_{in_0+j}\big|\calF_{(i-1)n_0+j}\big]\Big]\Big|  \to 0, \quad \mbox{a.s.}
\end{align}
Hence, we have that as $q\to \infty$
\begin{align}
\Big|\sum_{j=0}^{n_0-1} \sum_{i=1}^{q-1}b_n^{-1}\Big[U_{in_0+j}-\bbE\big[U_{in_0+j}\big|\calF_{(i-1)n_0+j}\big]\Big] \Big| &\leq \sum_{j=0}^{n_0-1}  b_n^{-1}\Big|\sum_{i=1}^{q-1}\Big[U_{in_0+j}-\bbE\big[U_{in_0+j}\big|\calF_{(i-1)n_0+j}\big]\Big] \Big|\\
\label{eqbndecreasing}
&\leq \sum_{j=0}^{n_0-1}   b_{(q-1)n_0+j}^{-1}\Big|\sum_{i=1}^{q-1}\Big[U_{in_0+j}-\bbE\big[U_{in_0+j}\big|\calF_{(i-1)n_0+j}\big]\Big] \Big|\\
&\to 0, \quad   \mbox{a.s.}
\end{align}
Here,~\eqref{eqbndecreasing} follows from the fact that $\{b_n\}_{n=1}^{\infty}$ is an increasing sequence.  This means that
\begin{align}
\label{eq237beauty}
\sum_{j=0}^{n_0-1} \sum_{i=1}^{q-1}b_n^{-1}\Big[U_{in_0+j}-\bbE\big[U_{in_0+j}\big|\calF_{(i-1)n_0+j}\big]\Big]\to 0,   \quad \mbox{a.s.}
\end{align}
Similarly, since for each fixed $j \in \{0,1,2,\ldots,r\}$ we have
\begin{align}
&\sum_{q=1}^{\infty} b_{qn_0+j}^{-2}\bbE\Big[\Big( U_{qn_0+j}-\bbE\big[U_{qn_0+j}\big|\calF_{(q-1)n_0+j}\big]\Big)^2\Big]\nonumber\\
\label{eqnewsup}
&=\bbE\Big[\sum_{q=1}^{\infty} b_{qn_0+j}^{-2}\bbE\Big[\Big( U_{qn_0+j}-\bbE\big[U_{qn_0+j}\big|\calF_{(q-1)n_0+j}\big]\Big)^2\Big|\calF_{(q-1)n_0+j}\Big]\Big] \\
&\leq \sum_{q=1}^{\infty} b_{qn_0+j}^{-2}\bbE\Big[\bbE\big[U_{qn_0+j}^2\big|\calF_{(q-1)n_0+j}\big]\Big] \\
&=\sum_{q=1}^{\infty} b_{qn_0+j}^{-2}\bbE\big[U_{qn_0+j}^2\big] \\
&\leq \sum_{n=1}^{\infty} b_n^{-2} \bbE[U_n^2]\\
&< \infty,
\end{align}
where~\eqref{eqnewsup} follows from the monotone convergence theorem~\cite{Billingsley}. Thus, as $q \to \infty$
\begin{align}
b_{qn_0+j}^{-1} \Big[U_{qn_0+j}-\bbE\big[U_{qn_0+j}\big|\calF_{(q-1)n_0+j} \big]\Big] \to 0, \quad \mbox{a.s.}
\end{align}  This holds because for a sequence of random variables $\{X_n\}_{n=1}^\infty$, if $\sum_{n=1}^\infty \bbE[ X_n^2]<\infty$,  then $X_n\to 0$ almost surely. This follows from a simple application of Chebyshev's inequality and the (first) Borel-Cantelli lemma. 
Furthermore, since the positive sequence $\{b_n\}_{n=1}^{\infty}$ is increasing, 
\begin{align}
b_n^{-1} \Big| U_{qn_0+j}-\bbE\big[U_{qn_0+j}\big|\calF_{(q-1)n_0+j}\big]\Big|\to 0, \quad \mbox{a.s.}
\end{align} for each fixed $j \in \{0,1,2,\ldots,r\}$. It follows that
\begin{align}
\Big|\sum_{j=0}^r b_n^{-1} \Big[U_{qn_0+j}-\bbE[U_{qn_0+j}\big|\calF_{(q-1)n_0+j}\big]\Big]\Big|&\leq \sum_{j=0}^r b_n^{-1} \Big| U_{qn_0+j}-\bbE\big[U_{qn_0+j}\big|\calF_{(q-1)n_0+j}\big]\Big|\\
&\leq \sum_{j=0}^{n_0-1} b_n^{-1} \Big| U_{qn_0+j}-\bbE\big[U_{qn_0+j}\big|\calF_{(q-1)n_0+j}\big]\Big| \to 0, \quad \mbox{a.s.}
\end{align}
Hence, we obtain
\begin{align}
\label{eq247beauty}
b_n^{-1} \sum_{j=0}^r \Big[U_{qn_0+j}-\bbE\big[U_{qn_0+j}\big|\calF_{(q-1)n_0+j}\big]\Big] \to 0, \quad \mbox{a.s.}
\end{align}
In addition, since $\sup_{n} b_n^{-1} \sum_{i=1}^n \bbE|U_i|< \infty$, we have
\begin{align}
f(n_0) b_n^{-1} \sum_{i=n_0}^n \bbE|U_i| &< f(n_0) b_n^{-1} \sum_{i=1}^n \bbE|U_i|\\
&\leq f(n_0) \sup_{n}b_n^{-1} \sum_{i=1}^n \bbE|U_i|\\
\label{beauty4}
&< \eps \sup_{n}b_n^{-1} \sum_{i=1}^n \bbE|U_i|. 
\end{align}
Combining~\eqref{eq23notenew},~\eqref{beauty1},~\eqref{eq237beauty},~\eqref{eq247beauty}, and~\eqref{beauty4} we obtain
\begin{align}
\limsup_{n\to \infty} \Big|b_n^{-1} \sum_{i=1}^n U_i\Big| < \eps \left(\sup_{n} b_n^{-1} \sum_{i=1}^n \bbE|U_i|\right), \quad \mbox{a.s.}
\end{align}
Take $\eps\to 0$, we have~\eqref{eqn:lemma12_concl},  
which completes the proof of Lemma~\ref{Heydelema}.
\end{IEEEproof}
\begin{lemma} \label{lem6a}
 Let $K$ and $K'$ be two positive constants  and let $\{\xi_n\}_{n=0}^{\infty}$ be a sequence adapted to filtration $\{\calF_n\}_{n=0}^{\infty}$ such that
\begin{align}
\bbE(\xi_{n+1}|\calF_n)&\geq \xi_n +K, \label{eqn:geK} \\
|\xi_{n+1}-\xi_n| &\leq K'. \label{eqn:leK'} 
\end{align} 
Let $\tau$ be a stopping time given by~\eqref{eqburnashevstopping}
for some $T \in \bbR$. Then, we have
\begin{align}
\bbP(\tau<\infty)=1. \label{eqn:Ptau1}
\end{align}
\end{lemma}
\begin{IEEEproof} [Proof of Lemma~\ref{lem6a}] \label{lem6a:proof}
We have
\begin{align}
\xi_n&=\xi_0+\sum_{i=1}^n (\xi_i-\xi_{i-1})\\
&=\xi_0 +\sum_{i=1}^n \bbE[\xi_i-\xi_{i-1}|\calF_{i-1}]+\sum_{i=1}^n[\xi_i-\xi_{i-1}-\bbE[\xi_i-\xi_{i-1}|\calF_{i-1}]]\\
\label{eq23newest2017}
&\geq \xi_0 + n K +\sum_{i=1}^n \beta_i,
\end{align}
where $
\beta_i=\xi_i-\xi_{i-1}-\bbE[\xi_i-\xi_{i-1}|\calF_{i-1}]$  
and~\eqref{eq23newest2017} follows from~\eqref{eqn:geK}.

Obviously, we have from~\eqref{eqn:leK'} that
\begin{align}
\label{eqlan0}
\bbE[\beta_i|\calF_{i-1}]&=0,\\
\bbE[\beta_i^2|\calF_{i-1}] &\leq \bbE[(\xi_i-\xi_{i-1})^2|\calF_{i-1}] \leq (K')^2, \quad \forall\, i\in\bbN.\label{eqlan0a}
\end{align}
 It is easy to see that the sequence $\{\beta_n\}_{n=1}^{\infty}$ is $*$-submixing adapted to $\{\calF_n\}_{n=1}^{\infty}$. Indeed, for $m, n\in \bbN$ we have
\begin{align}
\bbE[\beta_{n+m}|\calF_n]&=\bbE[\bbE[\beta_{n+m}|\calF_{n+m-1}]|\calF_m]\\
\label{eqlan1}
&=\bbE[0|\calF_m]=0.
\end{align}
Here,~\eqref{eqlan1} follows from~\eqref{eqlan0}. It follows from~\eqref{eqlan1} that
$
\bbE[\beta_{n+m}]=0.
$ 
Applying Lemma~\ref{Heydelema} for the $*$-submixing sequence $\{\beta_n\}_{n=1}^{\infty}$ adapted to $\{\calF_n\}_{n=1}^{\infty}$ and $b_n=n$ for all $n\in\bbN$, we obtain
\begin{align}
\lim_{n\to \infty} \frac{1}{n}\sum_{i=1}^n\beta_i = 0,\quad \mbox{a.s.}
\end{align}
It follows that
\begin{align}
\liminf_{n \to \infty} \frac{\xi_n}{n} \geq K>0, \quad \mbox{a.s.}
\end{align}
Hence, $\bbP(\tau <\infty)=1$. 
\end{IEEEproof}

\section{Proof of Lemma \ref{lem6}}\label{lem6:proof}
Before proving Lemma \ref{lem6}, we state some properties of $\xi_\tau$ and $\xi_{n\wedge\tau}$ in the following lemma.
\begin{lemma}\label{welldefined}
Under the same assumptions as Lemma~\ref{lem6a}, we have
\begin{itemize}
\item $\xi_{\tau}$, which is defined in Lemma \ref{lem6}, is a random variable, i.e., measurable with respect to~$\calF_{\infty}=\cup_{n=1}^{\infty} \calF_n=\lim_{n \to \infty} \calF_n$.
\item The following limiting statement holds:
\begin{align}
\label{dominated}
\lim_{n\to \infty} \bbE(|\xi_{n\wedge \tau}|)=\bbE(|\xi_{\tau}|).
\end{align}
\end{itemize}
\end{lemma}
\begin{IEEEproof}[Proof of Lemma~\ref{welldefined}] Applying Lemma~\ref{lem6a} with the identifications $K=\min\{K_1,K_2\}$  and $K'=K_3$ ($K_1$, $K_2$, and $K_3$ are defined in Lemma \ref{lem6}), we have $\bbP(\tau<\infty)=1$, hence $\xi_{n \wedge \tau}$ almost surely converges to $\xi_{\tau}$ as $n \to \infty$. This means that $\xi_{\tau}$ is a well-defined random variable since the limit of a sequence of Borel measurable functions is a Borel measurable function~\cite{Royden}. In addition, from Lemma~\ref{lem6a} we also have
$
\bbE(\tau) < \infty$. Now, observe that
\begin{align}
\xi_{n\wedge\tau}=\xi_0+\sum_{i=0}^{n\wedge\tau-1}(\xi_{i+1}-\xi_i).
\end{align}
Define
\begin{align}
\xi_{\rmm\rma\rmx}=|\xi_0|+\sum_{i=0}^{\infty}|\xi_{i+1}-\xi_i|\bone\{\tau>i\}.
\end{align}
It is easy to see that
$
|\xi_{n\wedge\tau}| \leq \xi_{\rmm\rma\rmx}
$  for all $n$.
Now, we also have 
\begin{align}
\bbE[\xi_{\rmm\rma\rmx}]&=\bbE\bigg[|\xi_0|+\sum_{i=0}^{\infty}|\xi_{i+1}-\xi_i|\bone\{\tau>i\}\bigg]\\
&=\bbE[|\xi_0|]+\bbE\bigg[\sum_{i=0}^{\infty}|\xi_{i+1}-\xi_i|\bone\{\tau>i\}\bigg]\\
&\leq \bbE[|\xi_0|]+ K_3 \sum_{i=0}^{\infty} \bbE\left[\bone\{\tau>i\}\right]\label{eqn:b6}\\
& = \bbE[|\xi_0|]+ K_3 \sum_{i=0}^{\infty} \bbP(\tau>i) \\
&=\bbE[|\xi_0|]+K_3 \bbE(\tau)\\
&< \infty,
\end{align}
 where \eqref{eqn:b6} follows from \eqref{eqbunarshev3} and Fatou's lemma~\cite{Billingsley}.
Since $|\xi_{n\wedge \tau}| \to |\xi_{\tau}|$ as $n \to \infty$,~\eqref{dominated} is obtained by the dominated convergence theorem~\cite{Billingsley}. 
\end{IEEEproof}
We are now ready to prove  Lemma~\ref{lem6}.
\begin{IEEEproof}[Proof of Lemma~\ref{lem6}]
The proof idea is based on~\cite{Burnashev75}. However, the submartingale construction is changed to account for all the points in Remark~\ref{rmk}. First, we choose $G\in\bbR$ such that the following equality holds:
\begin{align}
\label{eq76note}
G= \left|\frac{1}{K_2}-\frac{1}{K_1}\right|
\cdot\max_{-K_3\leq x\leq K_3}\left[\frac{x}{\exp(x)-1}\right].
\end{align}
It is easy to see that $G>0$. Now, define the following sequence
\begin{align}
\label{eqverykey}
\eta_n=\begin{cases}  -G+G \exp(\xi_n)+ \frac{\xi_n}{K_1}-n, &\quad\mbox{if}\quad  \xi_n <0,\\  \frac{\xi_n}{K_2} -n+ \frac{K_2}{K_1}-1, &\quad \mbox{if}\quad \xi_n \geq 0\end{cases},
\end{align} where $G$ are defined in~\eqref{eq76note}. First, we show that $(\eta_n,\calF_n)$ forms a submartingale, i.e.
\begin{align}
\label{submart}
\bbE(\eta_{n+1}|\calF_n) \geq \eta_n, \quad n=0,1,2,\ldots
\end{align}
We consider four different cases. 
\begin{itemize}
\item Case 1: For $\xi_n \geq 0$ and $\xi_{n+1} \geq 0$, we obtain
\begin{align}
\bbE[\eta_{n+1}|\calF_n]&=\bbE\left[\frac{\xi_{n+1}}{K_2}-(n+1)+\frac{K_2}{K_1}-1\Big|\calF_n\right]\\
\label{eq310310}
&\geq \frac{(\xi_n+K_2)}{K_2}-(n+1)+\frac{K_2}{K_1}-1\\
&=\frac{\xi_n}{K_2}-n + \frac{K_2}{K_1}-1\\
\label{eqkeynew3}
&=\eta_n.
\end{align}
Here,~\eqref{eq310310} follows from~\eqref{eqbunarshev2} and~\eqref{eqkeynew3} follows from~\eqref{eqverykey}. 
\item Case 2: For $\xi_n \geq 0$ and $\xi_{n+1}< 0$, we obtain
\begin{align}
\bbE[\eta_{n+1}|\calF_n]&=\bbE\left[-G+G \exp( \xi_{n+1})+\frac{\xi_{n+1}}{K_1}-(n+1)\Big|\calF_n\right]\\
\label{eq3202018}
&\geq -G+G \exp( \bbE[\xi_{n+1}|\calF_n])+\frac{1}{K_1}\bbE[\xi_{n+1}|\calF_n]-(n+1)\\
\label{sup1}
&\geq -G + G\exp(\xi_n +K_2)+\frac{1}{K_1}(\xi_n+K_2)-(n+1)\\
&\geq -G + G\exp(\xi_n) +\frac{1}{K_1}\xi_n -n +\frac{K_2}{K_1}-1\\
\label{eqkey93new}
&\geq \frac{\xi_n}{K_2}-n+\frac{K_2}{K_1}-1\\
\label{eqkey22017}
&= \eta_n.
\end{align}
Here,~\eqref{eq3202018} follows from the convexity of $\exp(x)$,~\eqref{sup1} follows from~\eqref{eqbunarshev2},~\eqref{eqkey93new} follows from the fact $0 \leq \xi_n \leq \xi_{n+1} +K_3 < K_3$ and~\eqref{eq76note}, and~\eqref{eqkey22017} follows from~\eqref{eqverykey}.
\item Case 3: For $\xi_n < 0, \xi_{n+1} \geq 0$, we have
\begin{align}
\label{eqkey2}
\bbE[\eta_{n+1}|\calF_n]&=\bbE\left[\frac{\xi_{n+1}}{K_2}-(n+1)+\frac{K_2}{K_1}-1\Big|\calF_n\right]\\
\label{eqvery2030}
&\geq \frac{(\xi_n+K_1)}{K_2}-(n+1)+\frac{K_2}{K_1}-1\\
&=\frac{\xi_n}{K_2}-n +\frac{K_1}{K_2}+\frac{K_2}{K_1}-2\\
\label{eq3000}
&\geq \frac{\xi_n}{K_2}-n\\
\label{eqnew2030}
&\geq -G+G\exp(\xi_n)+\frac{\xi_n}{K_1}-n\\
\label{final2030}
&=\eta_n.
\end{align}
Here,~\eqref{eqvery2030} follows from~\eqref{eqbunarshev1},~\eqref{eq3000} follows from the fact that
$
\frac{K_1}{K_2}+\frac{K_2}{K_1} \geq 2,
$ for any $K_1,K_2>0$, 
\eqref{eqnew2030} follows from the fact that $0>\xi_n\geq \xi_{n+1}-K_3 \geq -K_3$ and~\eqref{eq76note}, and~\eqref{final2030} follows from~\eqref{eqverykey}.

\item Case 4: For the case $\xi_n < 0, \xi_{n+1} < 0$, we have
\begin{align}
&\bbE\left[-G+G \exp( \xi_{n+1})+\frac{\xi_{n+1}}{K_1}-(n+1)\Big|\calF_n\right]\\
&=\bbE\left[\frac{\xi_{n+1}}{K_1}-\frac{\xi_n}{K_1}+ G \exp( \xi_{n+1})- G \exp(\xi_n)\Big|\calF_n\right]\nonumber\\
&\quad -G+ G\exp( \xi_n)+ \frac{\xi_n}{K_1}-(n+1)\\
\label{eq3281000}
&=\eta_n -1+\frac{1}{K_1} \bbE[\xi_{n+1}-\xi_n|\calF_n]+ G\exp(\xi_n)\bbE[\exp(\xi_{n+1}-\xi_n)-1|\calF_n]\\
\label{eq74moi}
&\geq \eta_n-1+ \frac{1}{K_1}\bbE[\xi_{n+1}-\xi_n|\calF_n]+G \exp(\xi_n)\bbE\left[\xi_{n+1}-\xi_n\big|\calF_n\right]\\
\label{new4000}
&\geq \eta_n +G  \exp(\xi_n)K_1\\
\label{eq76moi}
&\geq \eta_n.
\end{align}
Here,~\eqref{eq3281000} follows from~\eqref{eqverykey},~\eqref{eq74moi} follows from the fact that $\exp(x)-1 \geq x$, and~\eqref{new4000} follows from~\eqref{eqbunarshev1}.
\end{itemize}
Now, from~\eqref{eqverykey} we have
\begin{align}
\eta_{n\wedge \tau}&=\left[\frac{\xi_{n\wedge \tau}}{K_2}-n\wedge \tau +\frac{K_2}{K_1}-1\right]\bone\{\xi_{n\wedge \tau}\geq  0\}  \nonumber \\
&\quad + \left[ -G+G \exp( \xi_{n\wedge \tau})+\frac{\xi_{n\wedge \tau}}{K_1}-n\wedge \tau\right]\bone\{\xi_{n\wedge \tau} < 0\}\\
&=\left[\frac{K_2}{K_1}-1+\frac{\xi_{n\wedge \tau}}{K_2}\right]\bone\{\xi_{n\wedge \tau} \geq 0\} \nonumber \\
&\quad +\left[ -G+ G\exp( \xi_{n\wedge \tau})+\frac{\xi_{n\wedge \tau}}{K_1}\right]\bone\{\xi_{n\wedge \tau}< 0\} -(n\wedge \tau)\\
&\leq  \frac{\xi_{n\wedge \tau}}{K_2} \bone\{\xi_{n\wedge \tau} \geq 0\} + \frac{\xi_{n\wedge \tau}}{K_1}\bone\{\xi_{n\wedge \tau} < 0\}-(n\wedge \tau) \nonumber\\
&\quad +\left(\frac{K_2}{K_1}-1\right)\bone\{\xi_{n\wedge \tau}\geq  0\} + \left(-G+ G\exp( \xi_{n\wedge \tau})\right)\bone\{\xi_{n\wedge \tau}< 0\} \\
&\leq  \frac{\xi_{n\wedge \tau}}{K_2} \bone\{\xi_{n\wedge \tau} \geq 0\} + \frac{\xi_{n\wedge \tau}}{K_1}\bone\{\xi_{n\wedge \tau} < 0\}-(n\wedge \tau) \nonumber\\
\label{eqnearfinish}
&\quad +\left(\frac{K_2}{K_1}-1\right)\bone\{\xi_{n\wedge \tau}\geq  0\} \\
&\leq  \frac{\xi_{n\wedge \tau}}{K_2} \bone\{\xi_{n\wedge \tau} \geq 0\} + \frac{\xi_{n\wedge \tau}}{K_1}\bone\{\xi_{n\wedge \tau} < 0\}-(n\wedge \tau) +\left|\frac{K_2}{K_1}-1\right|\\
& \leq \frac{\xi_{n\wedge \tau}}{K_2}\bone\{\xi_{n\wedge \tau} \geq 0\}-(n\wedge \tau) + \left|\frac{K_2}{K_1}-1\right|\\
& \leq  \frac{|\xi_{n\wedge \tau}|}{K_2}-(n\wedge \tau)+ \left|\frac{K_2}{K_1}-1\right|.
\end{align}
Here,~\eqref{eqnearfinish} follows from the fact that $\left(-G+ G\exp( x)\right)\bone\{x< 0\}\leq 0$. Hence, from~\eqref{submart} we obtain
\begin{align}
\eta_0 &\leq \bbE(\eta_{n\wedge \tau})\\
&\leq \limsup_{n\to \infty} \bbE(\eta_{n \wedge\tau})\\
& \leq \limsup_{n \to \infty}\left[ \frac{|\xi_{n\wedge \tau}|}{K_2}-(n\wedge \tau)+ \left|\frac{K_2}{K_1}-1\right|\right]\\
\label{eqnewkey1}
&\leq  \frac{\bbE[|\xi_{\tau}|]}{K_2}-\bbE(\tau)+\left|\frac{K_2}{K_1}-1\right| \\
\label{eqnewkey2}
&\leq \frac{ |T|+K_3 }{K_2}-\bbE(\tau) + \left|\frac{K_2}{K_1}-1\right|. 
\end{align}
Here,~\eqref{eqnewkey1} follows from Lemma~\ref{welldefined} and~\eqref{eqnewkey2} follows from from the fact that $T\leq \xi_{\tau} \leq T+K_3$, so $|\xi_{\tau}| \leq \max\{|T|,|T+K_3|\}\leq |T|+K_3$. This means that
\begin{align}
\bbE(\tau) &\leq \frac{ |T|+K_3 }{K_2}-\eta_0+ \left|\frac{K_2}{K_1}-1\right|\\
\label{eqxizero}
&= \frac{ |T|+K_3 }{K_2}-\left(-G+G\exp(\xi_0)+\frac{\xi_0}{K_1}\right)\bone\{\xi_0<0\}-\left(\frac{\xi_0}{K_2}+\frac{K_2}{K_1}-1\right)\bone\{\xi_0\geq 0\} +\left|\frac{K_2}{K_1}-1\right|\\
&\leq \frac{ |T|+K_3 }{K_2}+G- \frac{\xi_0}{K_1}\bone\{\xi_0<0\}-\frac{\xi_0}{K_2}\bone\{\xi_0\geq 0\} +2\left|\frac{K_2}{K_1}-1\right|\\
\label{eqmuzero}
&=  -K_1^{-1}\xi_0 \bone\{\xi_0 <0\}-K_2^{-1} \xi_0\bone\{\xi_0\geq 0\}+K_2^{-1} |T|+f(K_1,K_2,K_3).
\end{align} Here,~\eqref{eqxizero} follows from~\eqref{eqverykey} and $f$ is some function of $K_1, K_2$, and $K_3$. This concludes the proof of Lemma~\ref{lem6}.
\end{IEEEproof}
\section{Proof of Lemma~\ref{lemmanew1}}\label{lemmanew1:proof}
Define the sequence of random variables
\begin{align}
\label{mode0}
\gamma_n=\xi_n+K_2n, \quad n=0,1,2,\ldots
\end{align}
It follows from~\eqref{bunarshev2} that
\begin{align}
\gamma_n &\in \calF_n,\quad n=0,1,2,\ldots,\\
\label{mode1}
\bbE(\gamma_{n+1}|\calF_n) &\leq \gamma_n,\quad \forall n \geq \tau_0.
\end{align}
Using Lemma~\ref{lem6a} with $K=K_1$, we know that
\begin{align}
\label{tau0finite}
\bbP(\tau_0 <\infty)=1.
\end{align}
Similarly, from~\eqref{tau0finite} and Lemma~\ref{lem6a}, we also obtain \eqref{eqn:Ptau1}, i.e., 
\begin{align}
\label{taufinite}
\bbP(\tau<\infty)=1
\end{align}
since the new initial value of the supermartingale starts at $\tau_0$, i.e.,
\begin{align}
|\xi_{\tau_0}| <|T_0|+K_3 < \infty.
\end{align}
It follows from~\eqref{tau0finite} and~\eqref{taufinite} that  
\begin{align}
\gamma_{\tau_0}&=\lim_{n\to \infty} \gamma_{\tau_0\wedge n} \in \calF_{\infty},\\
\label{welldefinetau}
\gamma_{\tau}&=\lim_{n\to \infty} \gamma_{\tau \wedge n} \in \calF_{\infty}.
\end{align} 
Now, since $\tau \geq \tau_0$, we have for any $n\geq 0$ that
\begin{align}
\label{mode2}
\gamma_{\tau \wedge n }=\gamma_{\tau_0\wedge n}+\sum_{k=0}^{n} (\gamma_{k+1}-\gamma_k)\bone\{\tau_0\wedge n \leq k < \tau \wedge n \}
\end{align}
Since $\tau_0$ is a stopping time of the filtration $\{\calF_n\}_{n=0}^{\infty}$, $\tau_0 \wedge n$ is also a stopping time of this filtration~\cite{Billingsley}. Define
\begin{align}
\calF_{\tau_0 \wedge n}=\{A \in \calF_{\infty}: A \cap \{\tau_0 \wedge n \leq k\} \in \calF_k, \; k=0,1,2,\ldots\}.
\end{align}
Now, for any $A \in \calF_{\tau_0 \wedge n}$, it is easy to see that
\begin{align}
\bone\{\tau_0\wedge n \leq k < \tau \wedge n\}\bone\{A\}&= \bone\{\{\tau_0 \wedge n \leq k\} \cap A\}-\bone\{\{\tau \wedge n \leq k\}\cap A \} \\
\label{eqverybeauty1}
&= \bone\{\{\tau_0 \wedge n \leq k\} \cap A\}-\bone\{\{\tau \wedge n \leq k\}\cap (\{\tau_0 \wedge n \leq k\}\cap A) \} \\
\label{eqverybeauty2}
&\in \calF_k.
\end{align}
Here,~\eqref{eqverybeauty1} follows from the assumption that $\tau\geq \tau_0$ and~\eqref{eqverybeauty2} follows from the fact that $\{\tau_0\wedge n \leq k\} \cap A \in \calF_k$ for any $A \in \calF_{\tau_0 \wedge n}$ and that $\tau\wedge n$ is a bounded stopping time of the same filtration $\{\calF_n\}_{n=0}^{\infty}$, i.e., $\{\tau \wedge n \leq k\}\in \calF_k$~\cite{Billingsley}.
 
It follows from~\eqref{mode1} that
\begin{align}
\bbE[(\gamma_{k+1}-\gamma_k)\bone\{\tau_0 \leq k < \tau\wedge n\}\bone\{A\}] \leq 0, \quad \forall k \geq \tau_0. \label{eqn:c14}
\end{align}
Therefore, we have
\begin{align}
\label{mode3}
&\bbE\bigg[\sum_{k=0}^n (\gamma_{k+1}-\gamma_k)\bone\{\tau_0\wedge n \leq k < \tau \wedge n \}\bone\{A\}\bigg]\\
&\quad=\bbE\bigg[\sum_{k=0}^n (\gamma_{k+1}-\gamma_k)\bone\{\tau_0\wedge n \leq k < \tau_0  \}\bone\{A\}\bigg]
+ \bbE\bigg[\sum_{k=0}^n (\gamma_{k+1}-\gamma_k)\bone\{\tau_0 \leq k < \tau \wedge n \}\bone\{A\}\bigg]\\
\label{complicated}
&\quad\leq \bbE\bigg[\sum_{k=0}^n (\gamma_{k+1}-\gamma_k)\bone\{\tau_0\wedge n \leq k < \tau_0  \}\bone\{A\}\bigg].
\end{align}
where \eqref{complicated} follows from \eqref{eqn:c14}. 
Now, observe that
\begin{align}
\label{newly10}
\sum_{k=0}^n (\gamma_{k+1}-\gamma_k)\bone\{\tau_0\wedge n \leq k < \tau_0  \}\bone\{A\}=\begin{cases}0,&\mbox{if}\quad \tau_0 \leq n\\ (\gamma_{n+1}-\gamma_n)\bone\{A\},&\mbox{if}\quad \tau_0 >n\end{cases}.
\end{align}
Hence, for all $n\geq 0$ we have from~\eqref{eqbunarshev3} and~\eqref{newly10} that 
\begin{align}
\label{complicated2}
\sum_{k=0}^n (\gamma_{k+1}-\gamma_k)\bone\{\tau_0\wedge n \leq k < \tau_0  \}\bone\{A\} \leq K_3.
\end{align}
From~\eqref{mode2},~\eqref{complicated}, and~\eqref{complicated2} we have
\begin{align}
\bbE(\gamma_{\tau \wedge n}\bone\{A\}) \leq \bbE(\gamma_{\tau_0 \wedge n} \bone\{A\})+K_3 \label{eqn:oneA}
\end{align} for any $A \in \calF_{\tau_0 \wedge n}$. Since $\calF_{\tau_0 \wedge n}$ is a $\sigma$-algebra~\cite{Billingsley}, therefore, by taking $A$ to be the entire sample space in~\eqref{eqn:oneA}, we obtain for any $n\geq 0$ that
\begin{align}
\label{mode4}
\bbE(\gamma_{\tau\wedge n} ) \leq \bbE(\gamma_{\tau_0} ) +K_3.
\end{align}
Now, we have for all $n\geq 0$ that
\begin{align}
\gamma_{\tau \wedge n}&=\gamma_0 +\sum_{k=0}^{\tau \wedge n} (\gamma_{k+1}-\gamma_k)\\
&\leq |\gamma_0| +\sum_{k=0}^{\tau \wedge n} |\gamma_{k+1}-\gamma_k|\\
&\leq |\gamma_0| +\sum_{k=0}^{\tau} |\gamma_{k+1}-\gamma_k|\\
&=|\gamma_0|+\sum_{k=0}^{\infty} |\gamma_{k+1}-\gamma_{k}|\bone\{\tau>k\}\\
\label{eqfinal2}
&= |\xi_0|+\sum_{k=0}^{\infty}|\xi_{k+1}-\xi_{k}+K_2|\bone\{\tau>k\}\\
&\leq |\xi_0|+\sum_{k=0}^{\infty}(|\xi_{k+1}-\xi_{k}|+K_2 )\bone\{\tau>k\}\\
\label{eqfinal1}
&\leq |\xi_0|+(K_3+K_2)\sum_{k=0}^{\infty}\bone\{\tau>k\}.
\end{align}
Here,~\eqref{eqfinal2} follows from~\eqref{mode0}, and~\eqref{eqfinal1} follows from~\eqref{eqbunarshev3}. Hence, we have for all $n\geq 0$ that
\begin{align}
\bbE|\gamma_{\tau \wedge n}| &\leq \bbE|\xi_0| +(K_2+K_3) \sum_{k=0}^{\infty} \bbP(\tau>k)\\
&=\bbE|\xi_0| +(K_2+K_3) \bbE(\tau)\\
\label{eqfinal3}
&< \infty.
\end{align}
Here,~\eqref{eqfinal3} follows from~\eqref{taufinite}. In addition, from~\eqref{tau0finite} and~\eqref{taufinite} we also have 
\begin{align}
\label{eqfinal6}
\lim_{n\to \infty} \gamma_{\tau \wedge n}=\gamma_{\tau}.
\end{align}
By the dominated convergence theorem~\cite{Billingsley} and from~\eqref{welldefinetau},~\eqref{eqfinal1},~\eqref{eqfinal3}, and~\eqref{eqfinal6} we have
\begin{align}
\label{newly2}
\lim_{n\to \infty}\bbE(\gamma_{\tau \wedge n})&=\bbE(\gamma_{\tau}).
\end{align}
From~\eqref{mode4} and~\eqref{newly2}  we obtain
\begin{align}
\label{mode5}
\bbE(\gamma_{\tau})\leq \bbE(\gamma_{\tau_0}) +K_3.
\end{align}
Combining~\eqref{mode0} and~\eqref{mode5} we have
\begin{align}
\bbE(\xi_{\tau}+\tau K_2) &\leq \bbE(\xi_{\tau_0}+\tau_0 K_2) +K_3.
\end{align}
Hence, we obtain 
\begin{align}
\bbE(\tau-\tau_0) &\leq \frac{\bbE(\xi_{\tau_0}-\xi_{\tau}) +K_3}{K_2}\\
\label{eqnewest}
&\leq \frac{T_0-T+3K_3}{K_2}.
\end{align}
Here,~\eqref{eqnewest} follows from the fact that $\xi_{\tau_0} \leq T_0+K_3$  and that $\xi_{\tau}\geq T-K_3$ if $T\leq T_0$. This proves \eqref{2020important} since $T_0\ge T$. 
\section{Proof of Lemma \ref{extralem}}\label{extralem:proof}
From~\eqref{eq35key2017} we have
\begin{align}
\liminf_{L\to \infty} \frac{-\ln \eps_L}{L\rho'_L}&\geq \liminf_{L\to \infty} \frac{-B}{p_{0,L} L \rho'_L}\Bigg[-\frac{L\rho'_L}{C}+\left[\frac{1}{C}-\frac{p_{0,L}}{B}+\frac{3(1-p_{0,L})}{2B^*}\right]Z_{0,L}+q_1(P_{Y|X})\Bigg]\\
\label{eq38}
&= \liminf_{L\to \infty} \frac{B}{Cp_{0,L}}-\frac{B}{p_{0,L}} \left[\frac{1}{C}-\frac{p_{0,L}}{B}+\frac{3(1-p_{0,L})}{2B^*}\right]\frac{Z_{0,L}}{L\rho'_L} -\frac{Bq_1(P_{Y|X})}{p_{0,L} L\rho'_L}. 
\end{align}
Note that
\begin{align}
\lim_{L\to \infty} p_{0,L}&=\lim_{L\to \infty} 1-\frac{1}{L}=1,\\
\lim_{L\to \infty} \frac{Z_{0,L}}{L\rho'_L}&=\lim_{L\to \infty}\frac{1}{L\rho'_L} \ln  \left(\frac{p_{0,L}}{1-p_{0,L}}\right)=\lim_{L\to \infty}\frac{\ln (L-1)}{L\rho'_L} =0\\
\label{app:D5}
\lim_{L\to \infty} \frac{1}{p_{0,L} L\rho'_L}&=\lim_{L\to \infty} \frac{1}{\sqrt{L} p_{0,L} (\rho'_L \sqrt{L})}=0.
\end{align}
It follows that
\begin{align}
\label{eq55key}
\liminf_{L\to \infty} \frac{-\ln \eps_L}{L\rho'_L} \geq \frac{B}{C}.
\end{align}
From~\eqref{eq55key} we have
\begin{align}
 \frac{-\ln \eps_L}{L\rho'_L}  \geq \frac{B}{2C}
\end{align} for $L$ sufficiently large. This is equivalent to
\begin{align}
\eps_L \leq \exp\left(-\frac{B}{2C}L\rho'_L\right)
\end{align} for $L$ sufficiently large.  Recall the definition of  $Z_{0,L}$ in~\eqref{eq145newest2017}. We have for $L$ sufficiently large that
\begin{align}
\exp(-A_L)-\eps_L \exp(-C_2) &=\exp(-Z_{0,L}/2)-\eps_L \exp(-C_2)\\
&=\frac{1}{\sqrt{L-1}} -\eps_L \exp(-C_2)\\
&\geq \frac{1}{\sqrt{L-1}} -\exp\left(-\frac{B}{2C}L\rho'_L\right) \exp(-C_2)\\
\label{eq194newest2017}
&\geq \frac{1}{\sqrt{L-1}} -\exp\left(-\frac{B}{2C}\sqrt{L}\right) \exp(-C_2)\\
& \ge \frac{1}{2 \sqrt{L-1} }.
\end{align}
Here,~\eqref{eq194newest2017} follows from the fact that $\sqrt{L}\rho'_L \to \infty $ as $L \to \infty$.  Therefore, we obtain for $L$ sufficiently large that
\begin{align}
p_{1,L}&\leq \frac{\exp(-Z_{0,L})-\eps_L \exp(-C_2)}{\exp(-A_L)-\eps_L \exp(-C_2)}\\
&= 2\sqrt{L-1} \left[\exp(-Z_{0,L})-\eps_L \exp(-C_2)\right]\\
&\leq 2\sqrt{L-1} \left[\frac{1}{L-1}-\eps_L \exp(-C_2)\right]\\
& \leq \frac{2}{\sqrt{L-1}}. \label{eqn:p1L}
\end{align}
Hence, we obtain from~\eqref{eq24key2017} that
\begin{align}
\label{eq31key2017}
p_{0,L}(1-p_{1,L})\bbE(W_L) &=-p_{0,L}\frac{\ln \eps_L}{B}+\frac{\ln(\exp(L(C-\rho'_L))-1)}{C}\nonumber\\
&\quad +\left[\frac{1}{C}-\frac{p_{0,L}}{B}+\frac{(1-p_{0,L})}{B^*}\right]Z_{0,L} +\frac{(1-p_{0,L})|A_L|}{B^*}+q_1(P_{Y|X})\\
&\leq -p_{0,L}\frac{\ln \eps_L}{B}+\frac{LC-L\rho'_L}{C}+\left[\frac{1}{C}-\frac{p_{0,L}}{B}+\frac{(1-p_{0,L})}{B^*}\right]Z_{0,L} \nonumber\\ 
&\quad +\frac{(1-p_{0,L})|A_L|}{B^*}+q_1(P_{Y|X})\label{eq206newest2017a}\\
\label{eq206newest2017}
&\leq L-p_{0,L}\frac{\ln \eps_L}{B}-\frac{L\rho'_L}{C}+\left[\frac{1}{C}-\frac{p_{0,L}}{B}+\frac{3(1-p_{0,L})}{2B^*}\right]Z_{0,L} +q_1(P_{Y|X})\\
\label{eq207newest2017}
&= L.
\end{align}
Here,~\eqref{eq206newest2017} follows from~\eqref{eq38keynew}  and~\eqref{eq207newest2017} follows from~\eqref{eq35key2017}. Hence, from \eqref{eqn:p1L},
\begin{align}
\bbE(W_L) &\leq \frac{L}{p_{0,L}(1-p_{1,L})} \leq \frac{L}{(1-1/L)(1-2/(\sqrt{L-1}))} 
\leq L + 3\sqrt{L},
\end{align} for $L$ sufficiently large.

\section{Proof of Lemma~\ref{lem5new}}\label{lem5new:proof}
\begin{IEEEproof}
The proof is based on a combination of  Burnashev's arguments in both~\cite{Burnashev1976} and~\cite{Burnashev80}. We can assume that $P_{Y|X}(y|x)>0$ for all $ x \in \calX, y\in \calY$, otherwise~\eqref{eq66new} trivially holds  since $B=\infty$. For each $i=1,2,\ldots,M$ and $y\in\calY$, define
\begin{align}
\label{eq70}
p_i&=\bbP(W=i|Y^n),  \\
p_i(y)&=\bbP(W=i|Y^n, Y_{n+1}=y),\label{eqn:piy} \\
p(y|W=i)&=\bbP(Y_{n+1}=y|Y^n, W=i),\\
p(y|W\neq i)&=\bbP(Y_{n+1}=y|Y^n, W\neq i),\\
\label{eq73}
p(y)&=\bbP(Y_{n+1}=y|Y^n).
\end{align}
We may assume without loss of generality that $p_i \neq 1$ for all $ i \in\calW=\{1,\ldots,M\}$. Otherwise, again the inequalities in~\eqref{eq66new} trivially hold. Using~\cite[Lemma 7]{Burnashev1976} and the definitions in~\eqref{eq70}--\eqref{eq73} we have
\begin{align}
\label{eq75new}
\bbE\left[\ln \calH(W|Y^n)-\ln \calH(W|Y^{n+1})\,\big|\,Y^n\right]&=\sum_{y \in \calY} p(y)\ln\bigg[\frac{-\sum_{i=1}^M p_i \ln p_i}{-\sum_{i=1}^M p_i(y) \ln p_i(y) }\bigg]\\
\label{eq76new}
&\leq \max_{i} \bigg\{\sum_{y \in \calY} p(y)\ln\left[\frac{- p_i \ln p_i}{-p_i(y) \ln p_i(y) }\right] \bigg\}  
\end{align}
Define 
\begin{equation}
F_i=\sum_{y \in \calY} p(y)\ln\bigg[\frac{ - p_i \ln p_i}{- p_i(y) \ln p_i(y) }\bigg] .
\end{equation}
It is easy to see that
\begin{align}
\label{eq77new}
p(y)&=p_i p(y|W=i)+(1-p_i) p(y|W\neq i),\\
\label{eq78new}
p_i(y)&=\frac{p_i p(y|W=i)}{p(y)},
\end{align}
and
\begin{align}
p(y|W=i)&=\bbP(Y_{n+1}=y|Y^n, W=i)\\
&=\sum_{x \in \calX} \bbP(X_{n+1}=x|W=i,Y^n) \bbP(Y_{n+1}=y|X_{n+1}=x, W=i,Y^n)\\
&=\sum_{x \in \calX} \bbP(X_{n+1}=x|W=i,Y^n) \bbP(Y_{n+1}=y|X_{n+1}=x) \label{eqn:use_mc}\\
&=\sum_{x \in \calX} \alpha_{ix} P_{Y|X}(y|x). \label{eqn:inv}
\end{align} Here, \eqref{eqn:use_mc} follows from the Markov chain $W-X_{n+1}-Y_{n+1}$ and  \eqref{eqn:inv} follows from the stationarity of the distribution $\bbP(Y_{n+1}=y|X_{n+1}=x)$ in $n$, which is derived from the stationarity of the distribution $\bbP(Y_{n+1}=y|X_{n+1}=x)$ in $n$.
Similarly, we have
\begin{align}
p(y|W\neq i)&=\bbP(Y_{n+1}=y|Y^n, W\neq i)\\
&=\sum_{x \in \calX} \bbP(X_{n+1}=x|W\neq i,Y^n) \bbP(Y_{n+1}=y|X_{n+1}=x, W\neq i,Y^n)\\
&=\sum_{x \in \calX} \bbP(X_{n+1}=x|W\neq i,Y^n) \bbP(Y_{n+1}=y|X_{n+1}=x)\\
& =\sum_{x \in \calX} \beta_{ix} P_{Y|X}(y|x).
\end{align}
It is easy to see that for each fixed message $i \in \calW = \{1,\ldots, M\}$ we have
\begin{align}
\label{eqkey}
\sum_{x \in \calX} \alpha_{ix}=\sum_{x\in \calX} \beta_{ix}=1, \quad \alpha_{ix}\geq 0, \beta_{ix} \geq 0.
\end{align}

Observe that $F_i$ is a function of variables $p_i, \{\alpha_{ix}\}$ and $\{\beta_{ix}\}$.  For the purpose of finding an upper bound on $\max_i \{F_i\}$ in~\eqref{eq76new}, we can consider only the constraints in~\eqref{eqkey} and find the maximization of $F_i$ over this convex set since other constraints that define the feasible set will only make $F_i$ smaller. With this consideration, let us consider find the maximization of $F_i$ over $\{\beta_{ix}\}$ with the assumption that $\sum_{x\in \calX} \beta_{ix}=1$ and $\beta_{ix} \geq 0$. Fix an arbitrary $x' \in \calX$, then we have $\beta_{ix'}=1-\sum_{x\in \calX\setminus\{x'\}}\beta_{ix}$. We readily obtain that the derivatives of $F_i$ for any $x\in \calX\setminus \{x'\}$ are
\begin{align}
\label{eq88}
\frac{\rmd^2 F_i}{\rmd\beta_{ix}^2}&=\frac{\partial^2 F_i}{\partial \beta_{ix}^2}+ \frac{\partial^2 F_i}{\partial \beta_{ix'}^2}-2 \frac{\partial^2 F_i}{\partial \beta_{ix} \partial \beta_{ix'}},\\
\frac{\partial^2 F_i}{\partial \beta_{ix} \partial \beta_{ix'}}&=(1-p_i)^2 \sum_{y \in \calY } \frac{\partial^2 F_i}{\partial p(y)^2} P_{Y|X}(y|x) P_{Y|X}(y|x'),\\
\label{eq90}
\frac{\partial^2 F_i}{\partial p(y)^2}&=\frac{1}{p(y)}\bigg[1-\left(\ln \frac{p(y)}{p_i p(y|W=i)}\right)^{-1}+\left(\ln \frac{p(y)}{p_i p(y|W=i)}\right)^{-2}\bigg] > 0.
\end{align}
Hence, from~\eqref{eq88} to~\eqref{eq90} we obtain
\begin{align}
\label{eq91new}
\frac{\rmd^2 F_i}{\rmd\beta_{ix}^2} =(1-p_i)^2 \sum_{y \in \calY } \frac{\partial^2 F_i}{\partial p(y)^2} \left(P_{Y|X}(y|x)-P_{Y|X}(y|x')\right)^2\geq 0, 
\end{align}
for any $x \in \calX\setminus\{x'\}$.

If for all $x\in \calX\setminus\{x'\}$ we have $D(P_{Y|X}(\cdot|x) \, \| \,  P_{Y|X}(\cdot|x'))=0$, it follows that 
\begin{align}
p(y|W=i)&=\sum_{x\in \calX} \alpha_{ix} P_{Y|X}(y|x)\\
&=\sum_{x\in \calX} \alpha_{ix} P_{Y|X}(y|x')\\
&=\sum_{x\in \calX \setminus \{x'\}} \alpha_{ix} P_{Y|X}(y|x')+\alpha_{ix'} P_{Y|X}(y|x') \\
&=(1-\alpha_{ix'})  P_{Y|X}(y|x')+ \alpha_{ix'} P_{Y|X}(y|x') \\
&=(1-\alpha_{ix'}) P_{Y|X}(y|x)+ \alpha_{ix'} P_{Y|X}(y|x) \\
&=P_{Y|X}(y|x), 
\end{align}
for any $i\in\calW$ and $y\in \calY$. 
In combination with the fact that the message is uniformly distributed on the message set $\calW$, we obtain
\begin{align}
p(y|W\neq i)= P_{Y|X}(y|x). \label{eqn:p_P}
\end{align}
Hence, it is easy to show that 
\begin{align}
p(y)= P_{Y|X}(y|x),\quad\mbox{and}\quad p_i(y)=p_i, \label{eqn:py}
\end{align}
for all $i\in\calW$  and $y\in \calY$. Therefore, we have
\begin{align}
\bbE\left[\ln \calH(W|Y^n)-\ln \calH(W|Y^{n+1})\,\big|\, Y^n\right]=0.
\end{align}

Now, we treat the remaining case where the relative entropy is positive. For any $x\in \calX$ there always exists an $x' \in \calX \setminus \{x\}$ such that $D(P_{Y|X}(\cdot|x)\,\|\, P_{Y|X}(\cdot|x'))>0$. By choosing $x'$ as a fixed symbol satisfying $D(P_{Y|X}(\cdot|x)\|P_{Y|X}(\cdot|x'))>0$,~\eqref{eq91new} becomes a strict  inequality. Therefore, $\beta_{ix}$ must be zero or one. Consequently, for all fixed $i\in\calW$, all the values of $\beta_{ix}$ for  all $ x\in \calX$ except for one are zero.  

Similarly, for any $x \in \calX \setminus \{x'\}$ such that $D(P_{Y|X}(\cdot|x) \,\|\, P_{Y|X}(\cdot|x'))>0$, we have
\begin{align}
\label{eq92key}
\frac{\partial^2 F_i}{\partial \alpha_{ix}^2}&=\sum_{y \in \calY } \left(P_{Y|X}(y|x)-P_{Y|X}(y|x')\right)^2 \frac{[p(y)-p_i p(y|W=i)]^2}{p(y) p^2(y|W=i)} \nn\\*
&\qquad\times \bigg[1-\left(\ln \frac{p(y)}{p_i p(y|W=i)}\right)^{-1}+\left(\ln \frac{p(y)}{p_i p(y|W=i)}\right)^{-2}\bigg] > 0.
\end{align}
Consequently, either $\alpha_{ix}=0$ or $\alpha_{ix}=1, x\in \calX$. 

From~\eqref{eq76new},~\eqref{eq77new}, and~\eqref{eq78new} together with above results, we obtain
\begin{align}
\label{eq93new}
\bbE\left[\calH(W|Y^n)-\calH(W|Y^{n+1})|Y^n\right]\leq \max\bigg\{0, \max_{x,x'}\max_{\eta} \bigg\{\sum_{y \in \calY} p(y) \ln \frac{\eta \ln \eta}{f(y)\ln f(y)}\bigg\}\bigg\},
\end{align}
where $\eta \in \{p_1,p_2,\ldots,p_M\}$, $(x, x') \in \calX^2$, and
\begin{align}
\label{eq94new}
p(y)&=\eta  P_{Y|X} (y|x)+(1-\eta) P_{Y|X} (y|x'), \\
\label{eq95new}
f(y)&=\eta \frac{P_{Y|X}(y|x)}{p(y)}.
\end{align}
 We see from~\eqref{eq94new} and~\eqref{eq95new} that
\begin{align}
\sum_{y \in \calY} p(y) \ln \frac{\eta\ln \eta}{f(y)\ln f(y)}=\sum_{y \in \calY} p(y) \ln \left[\frac{p^2(y)}{P_{Y|X}(y|x)P_{Y|X}(y|x')}\right]+\sum_{y \in \calY} p(y)\ln \left[\frac{P_{Y|X}(y|x')\ln \eta}{p(y)\ln f(y)}\right]. \label{eqn:e37}
\end{align}
Note that
\begin{align}
\frac{P_{Y|X}(y|x')}{p(y)}=\frac{1-f(y)}{1-\eta}.
\end{align}
It follows that
\begin{align}
\ln \left[\frac{P_{Y|X}(y|x')\ln \eta}{p(y)\ln f(y)}\right]&=\ln\left[\frac{(1-f(y))\ln \eta}{(1-\eta)\ln f(y)}  \right]\\
&=\left[\ln(1-f(y))-\ln(-\ln f(y))\right]-\left[\ln(1-\eta)-\ln(-\ln \eta)\right].
\end{align}
 
From~\eqref{eq95new}, we have
\begin{align}
\sum_{y \in \calY} p(y) f(y)=\sum_{y \in \calY} \eta P_{Y|X}(y|x)=\eta. \label{eqn:equals_eta}
\end{align} 
Combining with the fact that   $t\mapsto\ln(1-t)-\ln(-\ln t)$ is concave on $(0,1)$~\cite[pp.~424]{Burnashev80}, we obtain the following almost surely  
\begin{align}
\sum_{y \in \calY} p(y) \left[\ln(1-f(y))-\ln(-\ln f(y))\right] \leq \ln(1-\eta)-\ln(-\ln \eta). \label{eqn:app_jens}
\end{align} 
Note that $p(y)$ and $\eta$ are random because they    depend on $Y^n=(Y_1,\ldots, Y_n)$ which is   random (cf.~\eqref{eq70} and~\eqref{eqn:piy}).  The inequality in \eqref{eqn:app_jens}  means that
\begin{align}
\label{eq103new}
\sum_{y \in \calY} p(y)\ln \left[\frac{P_{Y|X}(y|x')\ln \eta}{p(y)\ln f(y)}\right] \leq 0.
\end{align}
 In addition, by Jensen's inequality  and the definition of $p(y)$ in \eqref{eq94new} (see the first inequality in Eqn.~(2.6) of~\cite{Burnashev80} for an analogous derivation), it holds that 
\begin{align}
 &p(y) \ln \left[\frac{p^2(y)}{P_{Y|X}(y|x)P_{Y|X}(y|x')}\right]\nn\\*
&\le \eta   P_{Y|X} (y|x) \ln \frac{P_{Y|X}(y|x)}{P_{Y|X}(y|x')} +(1-\eta) P_{Y|X} (y|x') \ln\frac{P_{Y|X}(y|x')}{P_{Y|X}(y|x)}
\end{align}
Hence, we obtain
\begin{align}
&\sum_{y \in \calY} p(y) \ln \left[\frac{p^2(y)}{P_{Y|X}(y|x)P_{Y|X}(y|x')}\right] \nn \\
&\le \eta  D(P_{Y|X} (\cdot|x)  \, \| \,   P_{Y|X}(\cdot |x' ) )  +(1-\eta)D(P_{Y|X} (\cdot|x' ) \|  P_{Y|X}(\cdot |x  ) )\\
\label{eq111new}
&\leq B\quad \mbox{a.s.}
\end{align}
From~\eqref{eq93new},~\eqref{eqn:e37},~\eqref{eq103new}, and~\eqref{eq111new} we obtain~\eqref{eq66new}, concluding the proof of Lemma~\ref{lem5new}.
\end{IEEEproof}

\section{Proof of Lemma \ref{lemimport}}\label{lemimport:proof}
From~\eqref{xnliminf} and definition of the limit inferior~\cite[Definition~3.16]{Rudin},  there exists a subsequence $\{x_{n_k}\}_{k=1}^{\infty}$ such that
$
\lim_{k\to \infty} x_{n_k}=0.
$ 
It then follows from~\eqref{eq219newest} that
$\limsup_{k\to\infty} y_{n_k}=0$. 
In addition, since  $\{y_n\}_{n\in \bbR_+}$ is non-negative,
$\lim_{k\to \infty} y_{n_k}=0$.
Since  $\{y_n\}_{n \in\bbR_+}$ is non-negative, again from the definition of the limit inferior,~\eqref{liminfy} holds.

\section{Proof of Lemma \ref{lemconverse}}\label{lemconverse:proof}
Assume, to the contrary of~\eqref{conversekey}, that
\begin{align*}
\limsup_{N\to \infty}-\frac{\ln \phi_N}{N}=+\infty.
\end{align*}
By definition of the limit superior~\cite[Definition~3.16]{Rudin}, there exists an increasing sequence of indices $\{N_k\}_{k=1}^{\infty}$ such that
\begin{align}
\label{eqkeypoint}
\lim_{k\to \infty}-\frac{\ln \phi_{N_k}}{N_k}=+\infty.
\end{align}
Then, we have
\begin{align}
&\limsup_{k\to \infty}\left[ -\frac{\ln \phi_{N_k}}{N_k\rho_{N_k}}-\frac{ \ln (N_k(C-\rho_{N_k})-\ln \phi_{N_k})}{N_k\rho_{N_k}} \right] \nonumber\\
&=\limsup_{k\to \infty}-\frac{\ln \phi_{N_k}}{N_k} \left[\frac{1}{\rho_{N_k}}-\frac{ \ln \left(\frac{CN_k-N_k\rho_{N_k}}{N_k}-\frac{\ln \phi_{N_k}}{N_k}\right)+\ln N_k}{-\frac{\ln \phi_{N_k}}{N_k}}\cdot \frac{1}{\sqrt{N_k} (\sqrt{N_k}\rho_{N_k})} \right]\\
\label{eq:G3}
&=\lim_{k\to \infty}-\frac{\ln \phi_{N_k}}{N_k} \Bigg[\frac{1}{\rho_{N_k}}-\frac{ \ln \left(\frac{CN_k-N_k\rho_{N_k}}{N_k}-\frac{\ln \phi_{N_k}}{N_k}\right)}{-\frac{\ln \phi_{N_k}}{N_k}}\cdot \frac{1}{\sqrt{N_k} (\sqrt{N_k}\rho_{N_k})}\Bigg]  -\frac{ \ln N_k}{\sqrt{N_k}}\cdot \frac{1}{ \sqrt{N_k}\rho_{N_k} }\\
\label{eq41key}
&=+\infty.
\end{align}
On the other hand, from~\eqref{eqkeypoint} we have
\begin{align}
-\frac{\ln \phi_{N_k}}{N_k} \geq 1
\end{align} for $k$ sufficiently large. This is equivalent to
\begin{align}
\label{eqn:eq416}
0\leq \phi_{N_k} \leq  \exp(-N_k),
\end{align} for $k$ sufficiently large. Hence, we obtain
\begin{align}
0\leq \limsup_{k\to \infty} \frac{\phi_{N_k}}{\rho_{N_k}}\leq \limsup_{k\to \infty} \frac{\exp(-N_k)}{\rho_{N_k}} = \limsup_{k\to \infty} \left[\exp(-N_k) \sqrt{N_k}\right] \left[\frac{1}{\sqrt{N_k}\rho_{N_k}}\right]
=0,
\end{align}
i.e.,  $\lim_{k\to \infty} {\phi_{N_k}}/{\rho_{N_k}}=0$. 
It follows that
\begin{align}
&\limsup_{k\to \infty}\left[ \frac{B }{C N_k\rho_{N_k}}+\frac{B \phi_{N_k}}{\rho_{N_k}}-\frac{B\phi_{N_k}}{C} +\frac{B}{C}+\frac{O(1)}{N_k\rho_{N_k}}\right] \nonumber \\
&\quad\leq \limsup_{k\to \infty} \frac{B }{C N_k\rho_{N_k}} +\limsup_{k\to \infty} \frac{B \phi_{N_k}}{\rho_{N_k}} -\liminf_{k \to \infty} \frac{B\phi_{N_k}}{C}+\frac{B}{C}+\limsup_{k\to \infty} \frac{O(1)}{N_k\rho_{N_k}}
\label{eq107key}
=\frac{B}{C}.
\end{align}
On the other hand, from~\eqref{eq37key} we have
\begin{align}
&-\frac{\ln \phi_{N_k}}{N_k\rho_{N_k}}-\frac{ \ln (N_k(C-\rho_{N,k})-\ln \phi_{N_k})}{N_k\rho_{N_k}} \nonumber\\
\label{conversekeybox}
&\quad\leq  \frac{B }{C N_k\rho_{N_k}}+\frac{B \phi_{N_k}}{\rho_{N_k}}-\frac{B\phi_{N_k}}{C} +\frac{B}{C}+\frac{O(1)}{N_k\rho_{N_k}}.
\end{align}
From~\eqref{eq41key},~\eqref{eqn:eq416},~\eqref{eq107key}, and~\eqref{conversekeybox} we obtain
\begin{align}
+ \infty &= \limsup_{k\to \infty}\left[ -\frac{\ln \phi_{N_k}}{N_k\rho_{N_k}}-\frac{ \ln (N_k(C-\rho_{N,k})-\ln \phi_{N_k})}{N_k\rho_{N_k}} \right] \\
&\leq \limsup_{k\to \infty}\left[ \frac{B }{C N_k\rho_{N_k}}+\frac{B \phi_{N_k}}{\rho_{N_k}}-\frac{B\phi_{N_k}}{C} +\frac{B}{C}+\frac{O(1)}{N_k\rho_{N_k}}\right] \leq \frac{B}{C}.
\end{align}
This is a contradiction, concluding the proof of  Lemma \ref{lemimport}. 

\paragraph*{Acknowledgements}
The authors would like to sincerely thank Dr.\ Mladen Kova\v{c}evi\'{c} (National University of Singapore) for useful comments that helped to greatly improve the manuscript.
\bibliographystyle{unsrt}
\bibliography{isitbib}
\end{document}